\documentclass[iop]{emulateapj-rtx4}
\usepackage{amsmath}
\usepackage[usenames,dvipsnames]{xcolor}

\usepackage[normalem]{ulem}
\usepackage{cancel}

\newcommand{\MJup}{M_{\rm Jup}}
\newcommand{\Msolar}{{\rm M_{\odot}}}   
\newcommand{\Rp}{R_{\rm p}}

\newcommand{\Mstar}{M_{\star}}

\newcommand{\torb}{t_{\rm orb}}
\newcommand{\tdrift}{t_{\rm drift}}
\newcommand{\icarus}{Icarus}
\newcommand{\St}{{\rm St}}

\shorttitle{Long-lasting dust rings}
\shortauthors{Meru et al.}

\begin{document}

\title{Long-lasting dust rings in gas-rich disks: sculpting by single and multiple planets}

\author{Farzana Meru\altaffilmark{1,2}}
\author{Sascha P. Quanz\altaffilmark{1}}
\author{Maddalena Reggiani\altaffilmark{1}}
\author{Cl\'ement Baruteau\altaffilmark{3}}
\author{Jaime E. Pineda\altaffilmark{1}}

\affil{$^1$Institut f\"ur Astronomie, ETH Z\"urich, Wolfgang-Pauli-Strasse 27, 8093 Z\"urich, Switzerland}
\email{farzana.meru@ast.cam.ac.uk}
\affil{$^2$Institute of Astronomy, University of Cambridge, Madingley Road, Cambridge, CB3 0HA, United Kingdom}
\affil{$^3$
Institut de Recherche en Astrophysique et Plan\'etologie, CNRS / Universit\'e de Toulouse / UPS-OMP, 14 avenue Edouard Belin, 31400 Toulouse, France}

\begin{abstract}

We propose a mechanism by which dust rings in protoplanetary disks can form \emph{and} be long-lasting compared to gas rings.  This involves the existence of a pressure maximum which traps dust either in between two gap-opening planets or at the outermost gap edge of a single or multiple planet system, \emph{combined with} the decoupling of large dust particles from the gas.  We perform 2D gas hydrodynamical simulations of disks with one and two giant planets which may open deep or partial gaps.  A gas ring forms in between two planets such that the surface mass density is higher than on either side of it.  This ring is a region of pressure maximum where we expect large grains, which are marginally coupled to the gas and would otherwise be subject to radial drift, to collect.  Such a pressure maximum also occurs at the outermost gap edge in a disk with one or more planets.  We infer the dust evolution in these regions as the gas disk evolves, to understand the longer term behavior of the resulting dust rings.  Over time the gas surface density in the ring(s) decreases, which may cause the larger trapped particles to decouple.  Consequently, these particles are expected to stay in ring structure(s) longer than the gas.  For a Minimum Mass Solar Nebula model, we expect that millimeter and centimeter-sized grains in the outer O(10) au would most likely undergo this trapping and decoupling process.

\end{abstract}


\section{Introduction}

Recent advances in observing facilities and techniques have provided evidence via spatially-resolved observations that a large variety of structures exist in protoplanetary disks including spirals \citep[e.g.][]{Boccaletti_spirals_HD100546,Avenhaus_HD142527}, azimuthal asymmetries \citep[e.g.][]{Isella_LkHalpha_multiplanet,Nienke_dust_trap,Perez_TD_asymmetries}, eccentric inner holes \citep{Thalmann_LkCa15_ecc} and ring-like structures. These structures may be signposts of one or more nascent or developed planets interacting with their parent disk. Ring-like structures have been observed in HD169142 and HD100546 but at different wavelengths: HD169142 shows a bright ring in scattered light data \citep{Quanz_doughnut} and some evidence of a ring-like feature at 7mm wavelengths \citep{Osorio_HD169142_7mm}, while HD100546 appears to be continuous in the scattered light \citep{Avenhaus_HD100546_cavity} but has a ring-like structure in the millimeter dust emission \citep{Pineda_HD100546,Walsh_HD100546}.  Interestingly both of these disks may even host multiple planets (see \citealp{Reggiani_HD169142,Biller_HD169142,Osorio_HD169142_7mm} for HD169142 and \citealp{Brittain_HD100546_candidate,Brittain_HD100546_candidate2,HD100546b,Currie_HD100546b_confirm,Quanz_HD100546_confirmation} for HD100546) and the ring structures appear to be located in between the two planets.  Furthermore, \cite{Walsh_HD100546} also showed an additional ring structure exterior to the outermost planet.  The very recent release of the HL Tau\footnote{http://www.eso.org/public/news/eso1436/} image using ALMA configured almost in its end state has shown that such ring structures can occur even in the very young disk phase.

These are currently the best examples of ring-like structures in gas-rich disks.  Upcoming facilities providing high-resolution and high-sensitivity observations, e.g. ALMA, SPHERE and SKA, may detect more ring-like structures in the future.

How might such ring structures form?  One mechanism, as suggested through observations, may be the presence of multiple planets.  \cite{Bryden_dust_multiplanet} performed numerical simulations of two planets embedded in a disk and found that a ring of gas forms in between them.  Eventually (after $\gtrsim{\rm O}(100)$ orbits) the ring disappeared as the planets grew in mass.

Multiple planet simulations in gas-rich protoplanetary disks have often been performed in the context of understanding planet migration and the capture of planets into mean motion resonances.  More recently a number of studies have focussed on the formation of common gaps with a view to describe the presence of large inner holes in transition disks \citep[e.g.][]{Pierens_Nelson_2planets,Zhu_multiplanet,DodsonRobinson_multiplanet,Isella_LkHalpha_multiplanet} or have considered multiple planets to potentially explain azimuthally asymmetric disk structures \citep[e.g.][]{Isella_LkHalpha_multiplanet,Pinilla_multiplanet}.  With regards to the former it must be noted that if two planets are maintained sufficiently far apart they may not form a common gap.

Motivated by recently observed ring-like features we perform hydrodynamical simulations to explore the persistence of gas rings in protoplanetary disks that harbor both single and multiple planets.  To link the gas hydrodynamical simulations with the observations, the connection between the gas and dust needs to be made.   We use analytics to infer how the dust in the disk would be affected by the gas ring(s), and in particular, what this means for the longer term dust disk structure, and hence observations.  In Section~\ref{sec:method} we outline our method and simulations performed, and illustrate our results in Section~\ref{sec:results}.  We then discuss the results and conclude in Sections~\ref{sec:disc} and~\ref{sec:conc}, respectively.

\section{Method \& Simulations}
\label{sec:method}

\begin{table*}
\centering
\begin{tabular}{lll}
 \tableline
   \hline
   Parameter & {\sc fargo} code units & Application to MMSN disk\\
    \tableline
    Stellar mass, $\Mstar$ & 1 & $1 \Msolar$\\
    Inner radius, $R_{\rm min}$ & 0.4 & 6 au\\
    Outer radius, $R_{\rm max}$ & 4 & 60 au\\
    Initial gas surface density, $\Sigma_o (R)$ & $3.7 \times 10^{-6} R^{-3/2}$ & $1700 (R/1au)^{-3/2} \rm ~g~cm^{-2}$ \\
    Temperature, $T$ & $2.5 \times 10^{-3} R^{-1}$ & $42 (R/15 \rm au)^{-1}$ K\\
    Aspect ratio, $H/R$ & 0.05 (constant) & 0.05 (constant)\\
    Viscosity parameter & $\alpha = 4 \times 10^{-3} R^{-0.5}$ & $\alpha = 4 \times 10^{-3} (R/15 {\rm au})^{-0.5}$\\
   \hline
 \end{tabular}
  \caption{Disk properties in code units and applying this to the MMSN disk.}
\label{tab:ref_disc}
\end{table*}

\begin{table*}
\centering
{\scriptsize
\begin{tabular}{lllllll}
  \tableline
   \hline
    Simulation & Type of & Planet mass(es) & Planet mass(es) & Planet radial location(s) & Planet radial location(s) & Planet\\
    & simulation & [code units] & [applied to MMSN disk] & [code units] & [applied to MMSN disk] & period ratio\\
    \tableline
    1 & Single planet & $1 \times 10^{-3} \Mstar$ & $1 \MJup$ & $R_{\rm p} = 1$ & $R_{\rm p} = 15$~au & --\\
    2 & Single planet & $2 \times 10^{-4} \Mstar$ & $0.2 \MJup$ & $R_{\rm p} = 1$ & $R_{\rm p} = 15$~au & --\\
    3 & Multiplanet & $1 \times 10^{-3} \Mstar$ & $1 \MJup$ & $R_{\rm p, in} = 1$, $R_{\rm p, out} = 2$ & $R_{\rm p, in} = 15$~au, $R_{\rm p, out} = 30$~au & 2.8\\
    4 & Multiplanet & $2 \times 10^{-4} \Mstar$ & $0.2 \MJup$ & $R_{\rm p, in} = 1$, $R_{\rm p, out} = 2$ & $R_{\rm p, in} = 15$~au, $R_{\rm p, out} = 30$~au & 2.8\\
  \hline
 \end{tabular}
}
  \caption{Planet properties in the simulations performed, both in code units and applying this to a MMSN disk.  The subscripts 'in' and 'out' refer to the inner and outer planets, respectively.}
\label{tab:ref_sim}
\end{table*}

We carry out 2D hydrodynamical simulations using the grid-based code {\sc fargo} \citep{Masset_FARGO}.  We model the disk using an isothermal equation of state, i.e. the disk temperature profile does not change over time, so that the results are scale-free.  While the results are general such that they can be applied to many disk properties, we also discuss the results specifically in the context of a  Minimum Mass Solar Nebula (MMSN) disk model in Section~\ref{sec:results_dust}.  Table~\ref{tab:ref_disc} gives the disk parameters in code units and the conversion to a MMSN model.  The code unit for the mass is given by $1\Msolar$ while that for the length and time is given by the radial location and orbital timescale of the planet (or the innermost planet in the multiple planet case), respectively.

Using the same disk setup throughout, we simulate the disk evolution under the presence of a single and two planets.  We perform two simulations with planet to star mass ratios, $q = 1 \times 10^{-3}$ and $2 \times 10^{-4}$, that are expected to open deep and partial gaps, respectively.  We perform two further simulations: firstly including two planets with $q = 1 \times 10^{-3}$ and secondly including two planets with $q = 2 \times 10^{-4}$ (see Table~\ref{tab:ref_sim}).

In both the single and multiplanet simulations the planets' masses are increased linearly from zero to the above-mentioned masses over the first orbit (at $R = 1$ in code units) of the simulation.  We also confirm that the results do not change if the mass is increased over 10 orbits.  For simplicity the planets' orbits are fixed, though we also perform test simulations with migrating planets and find the qualitative results to be unchanged (see Section~\ref{sec:further_cond}).

In the multiplanet case the planets are held at $R = 1$ and $2$, i.e. far enough apart such that their gaps are not expected to overlap.  The gap half-width in the gas is $\approx 2.5 R_{\rm H}$ \citep{Masset_2.5RH} where $R_{\rm H} = \Rp (q/3)^{1/3}$ is the Hill radius and $\Rp$ is the planet's radial location.  Thus the planets' gaps will not overlap if $(R_{\rm p, in} + 2.5 R_{\rm H,in}) < (R_{\rm p, out} - 2.5 R_{\rm H,out})$ where the subscripts 'in' and 'out' refer to the inner and outer planets, respectively.  Equivalently, the gaps will not overlap if the planet period ratio satisfies

\begin{equation}
\frac{P_{\rm out}}{P_{\rm in}} \gtrsim \left ( \frac{1 + 2.5(q_{\rm out}/3)^{(1/3)}}{1 - 2.5(q_{\rm in}/3)^{(1/3)}} \right )^{3/2},
\label{eq:ring}
\end{equation}
where the equality is $\approx 1.7$ and $\approx 1.4$ for two planets with $q = 1 \times 10^{-3}$ and $q = 2 \times 10^{-4}$, respectively.  The period ratio of these planets is $\approx 2.8$ which is within the observed range of ratios in multiplanet systems.

We model the disk using 277 and 800 radial and azimuthal cells, respectively, and ensure that the conceptual arguments presented in this paper still hold with twice the resolution in each direction.  With the grid resolution used the Hill radii of the planets with $q = 1 \times 10^{-3}$ are resolved in the radial direction using 5 and 11 grid cells for the inner and outer planets, respectively, while that of the planets with $q = 2 \times 10^{-4}$ are resolved using 3 and 6 grid cells, respectively.  In the azimuthal direction the planets' Hill radii are modelled using 9 and 5 cells for the $q = 1 \times 10^{-3}$ and $q = 2 \times 10^{-4}$ planets, respectively.  The simulations are evolved using open boundary conditions. 

The simulations performed are gas hydrodynamic simulations only.  To provide the link with the longer-term dust behavior in the disk we determine how well coupled the particles are to the gas and whether they are likely to be subject to radial drift, by calculating the Stokes numbers of particles of various sizes.  This is given by \citep{Garaud_vel_pdf}

\begin{equation}
\St = t_{\rm s} \Omega = \frac{\sqrt{2 \pi} s \rho_s}{\Sigma},
\end{equation}
where $t_{\rm s}$ is the stopping timescale due to gas drag, $s$ is the particle size and $\rho_s$ is the solid density assumed to be $1 \rm g cm^{-3}$ for $\rm SiO_2$ particles.  We then infer their likely behavior as the gas disk evolves over time.  While this expression for the Stokes number is for particles in the Epstein regime (i.e. for particle sizes $\lesssim (9/4) \lambda_{\rm mfp}$ where $\lambda_{\rm mfp}$ is the mean free path of the gas molecules, and for which collision velocities are subsonic), the mechanism proposed is general and would still be valid if the particles are in the Stokes regime (i.e. for particle sizes $\gtrsim (9/4) \lambda_{\rm mfp}$).

\section{Results}
\label{sec:results}

\subsection{Hydrodynamical results}

\begin{figure*}
\centering
\includegraphics[width=0.54\columnwidth,angle=90]{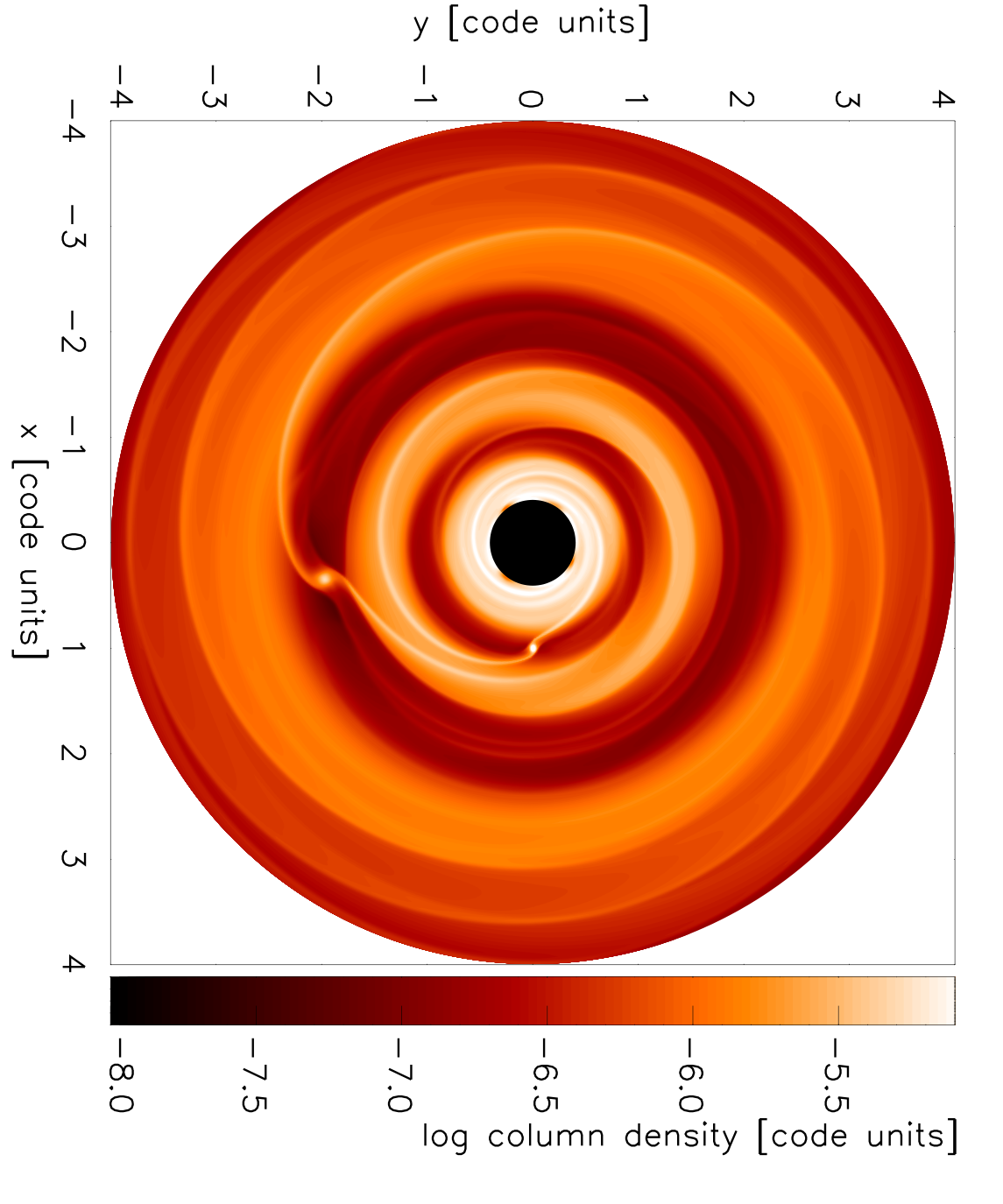}
\includegraphics[width=0.54\columnwidth,angle=90]{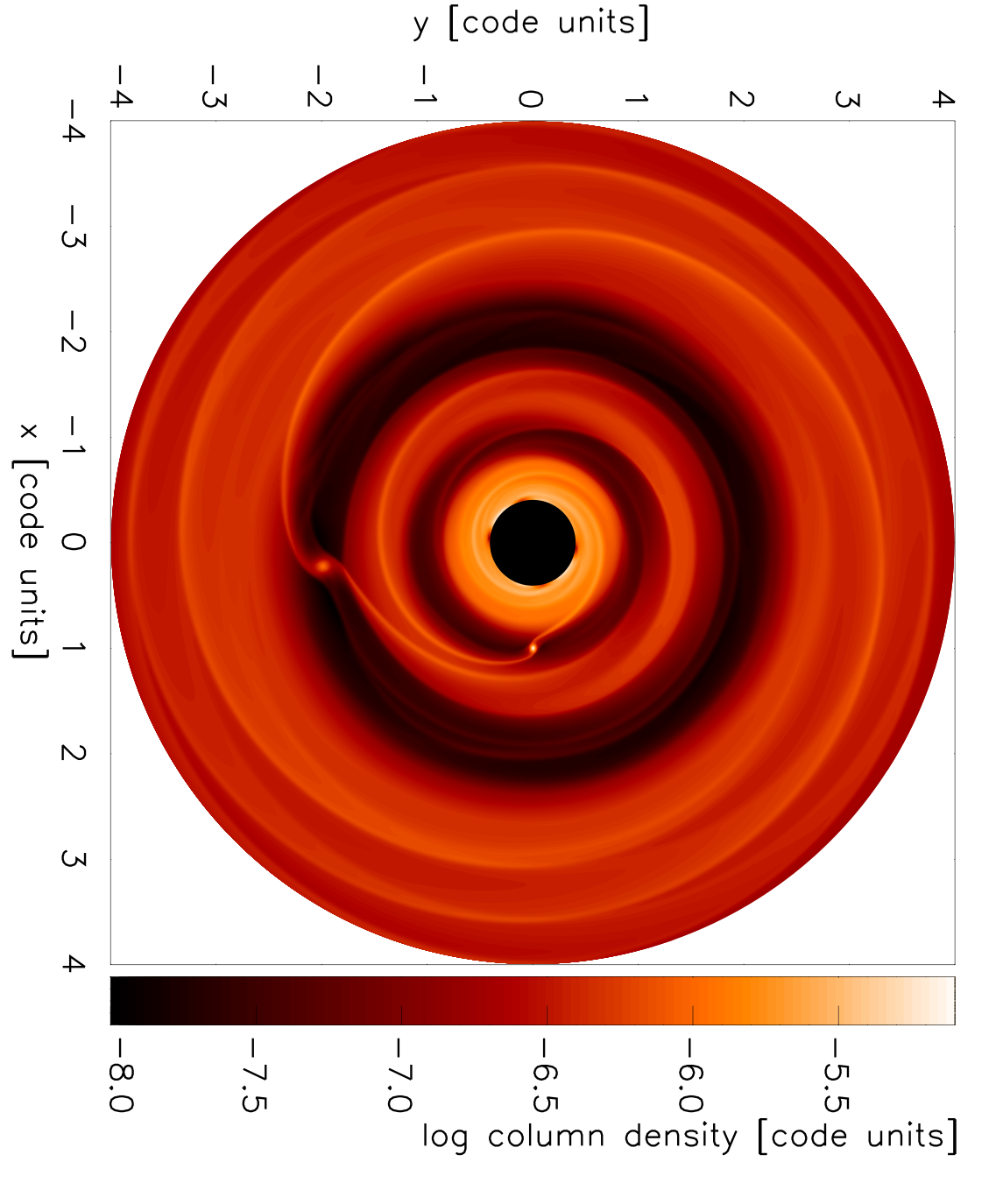}
\includegraphics[width=0.77\columnwidth]{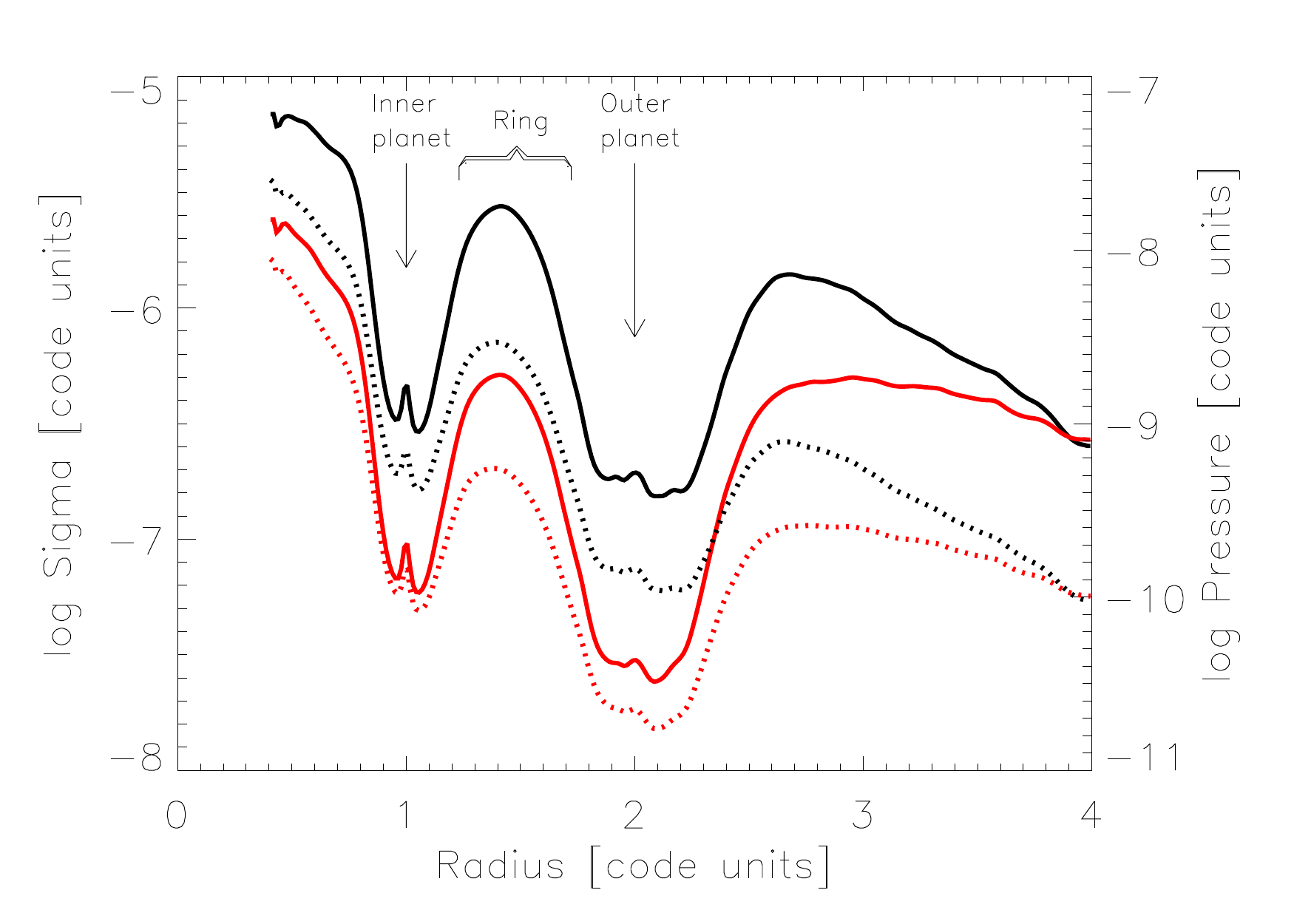}
\includegraphics[width=0.54\columnwidth,angle=90]{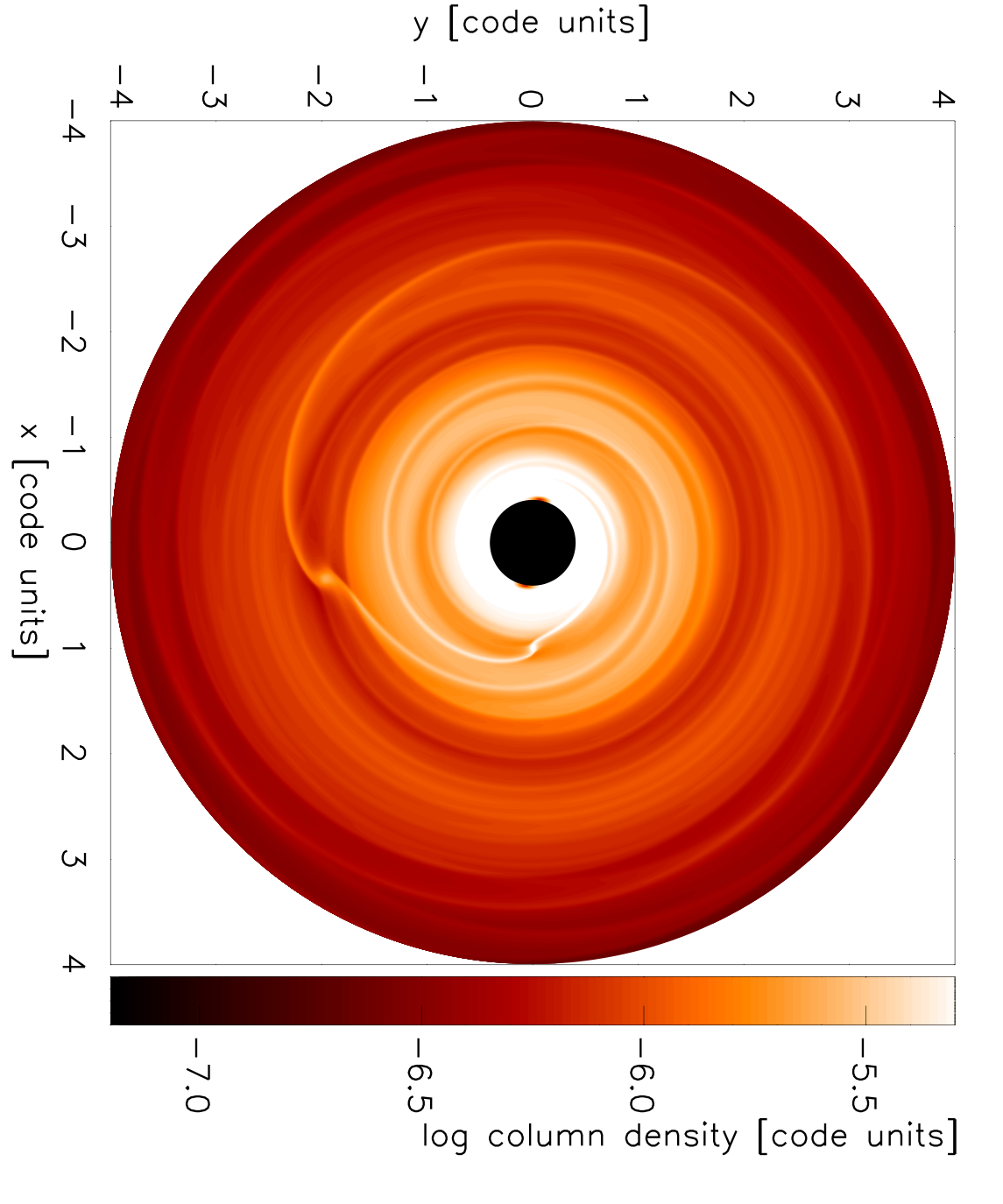}
\includegraphics[width=0.54\columnwidth,angle=90]{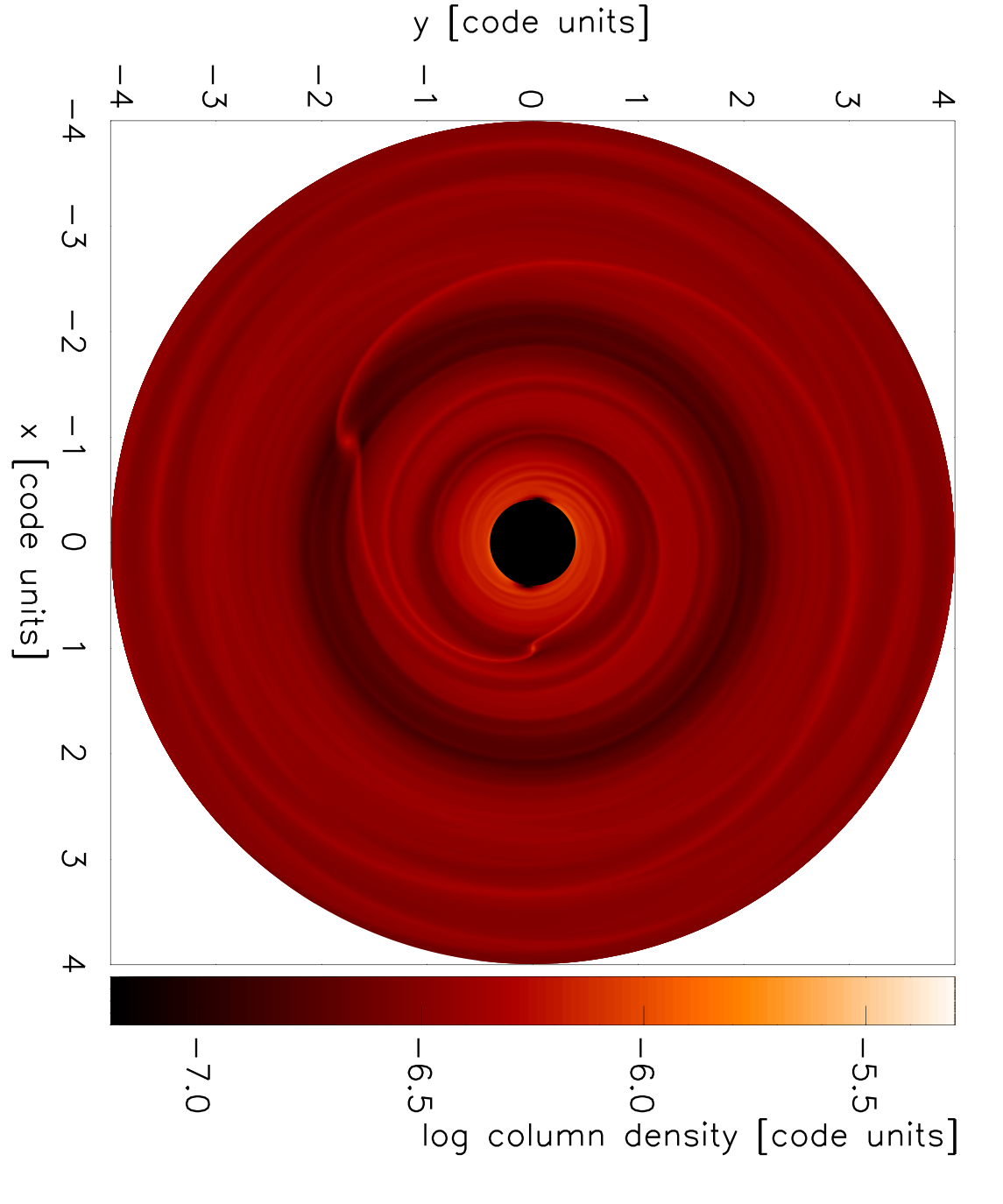}
\includegraphics[width=0.77\columnwidth]{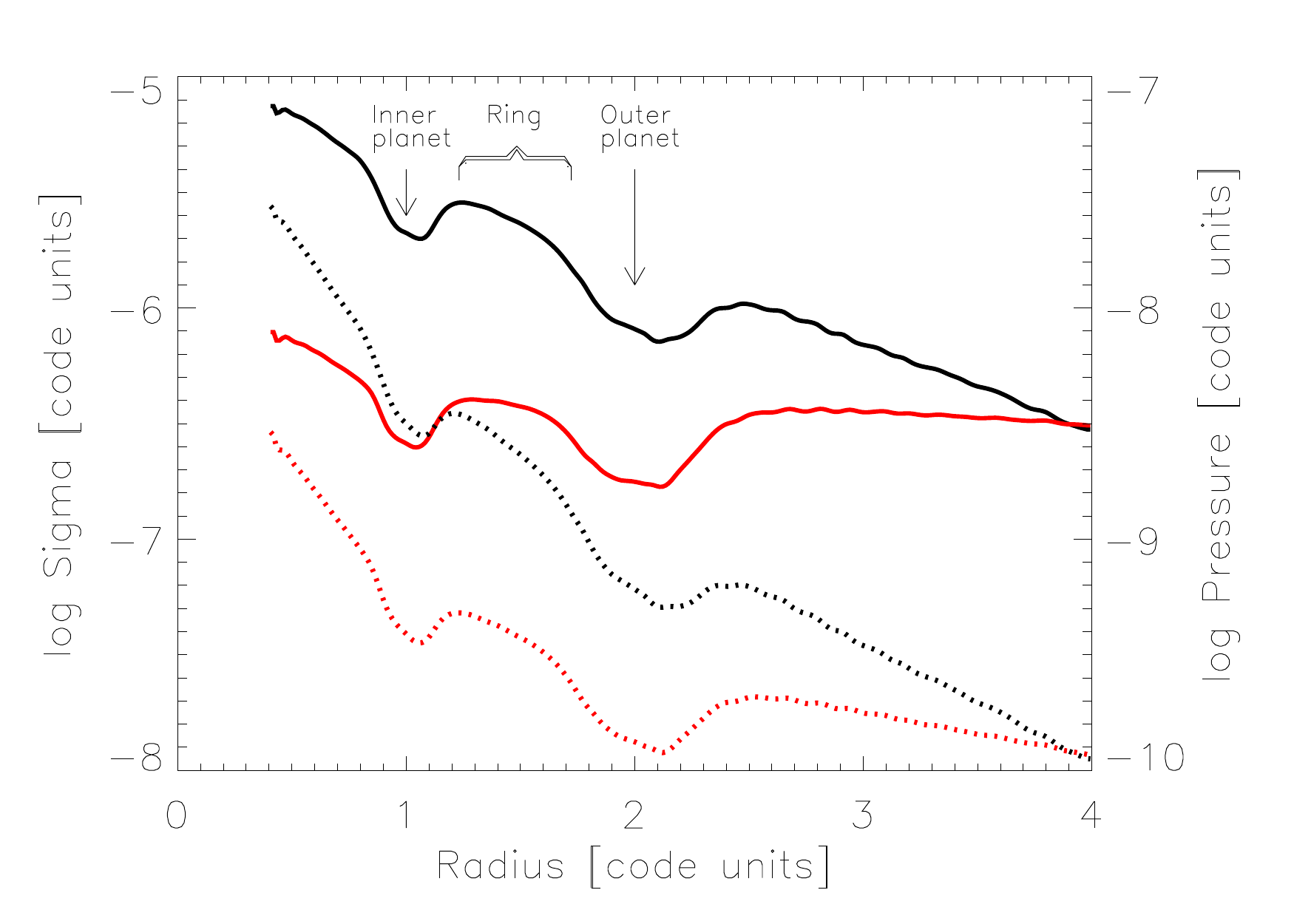}
\caption{Surface mass density rendered images of the gas disks and the azimuthally averaged surface mass density and pressure profiles of the disk with two $1 \MJup$ (top row) and two $0.2 \MJup$ (bottom row) planets.  The images and profiles are at an early time (t = 500 orbits) and later time (t = 5,000 and 50,000 orbits for the $1 \MJup$ and $0.2 \MJup$ simulations, respectively).  Left column: At an earlier time a ring of material collects in between the two planets irrespective of whether the planets have opened deep (top panel) or partial (bottom panel) gaps.  The surface mass density at the outermost gap edge is also high.  Middle column: At a later time the surface density in between the two planets and at the outermost gap edge decreases.  Right column: The surface mass density (solid lines) and pressure (dotted lines) decrease between the earlier (black lines) and late times (red lines).  A pressure maximum exists in between the two planets and at the outermost gap edge, where particles can be trapped and reside.  The decrease in the gas surface mass density causes an increase in the Stokes numbers of the particles and causes them to start decoupling from the gas.}
\label{fig:discs}
\end{figure*}

\begin{figure*}
\centering
\includegraphics[width=0.54\columnwidth,angle=90]{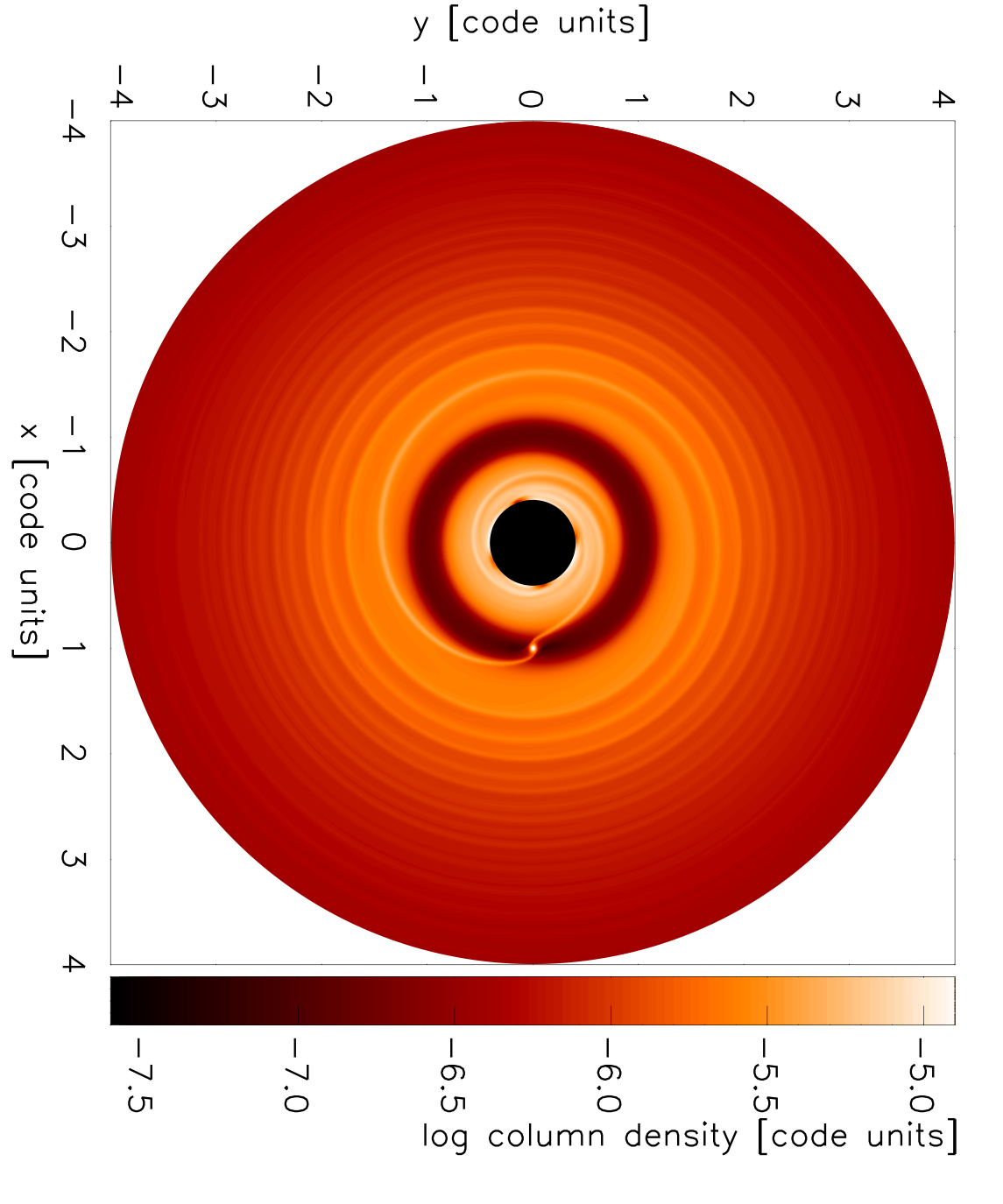}
\includegraphics[width=0.54\columnwidth,angle=90]{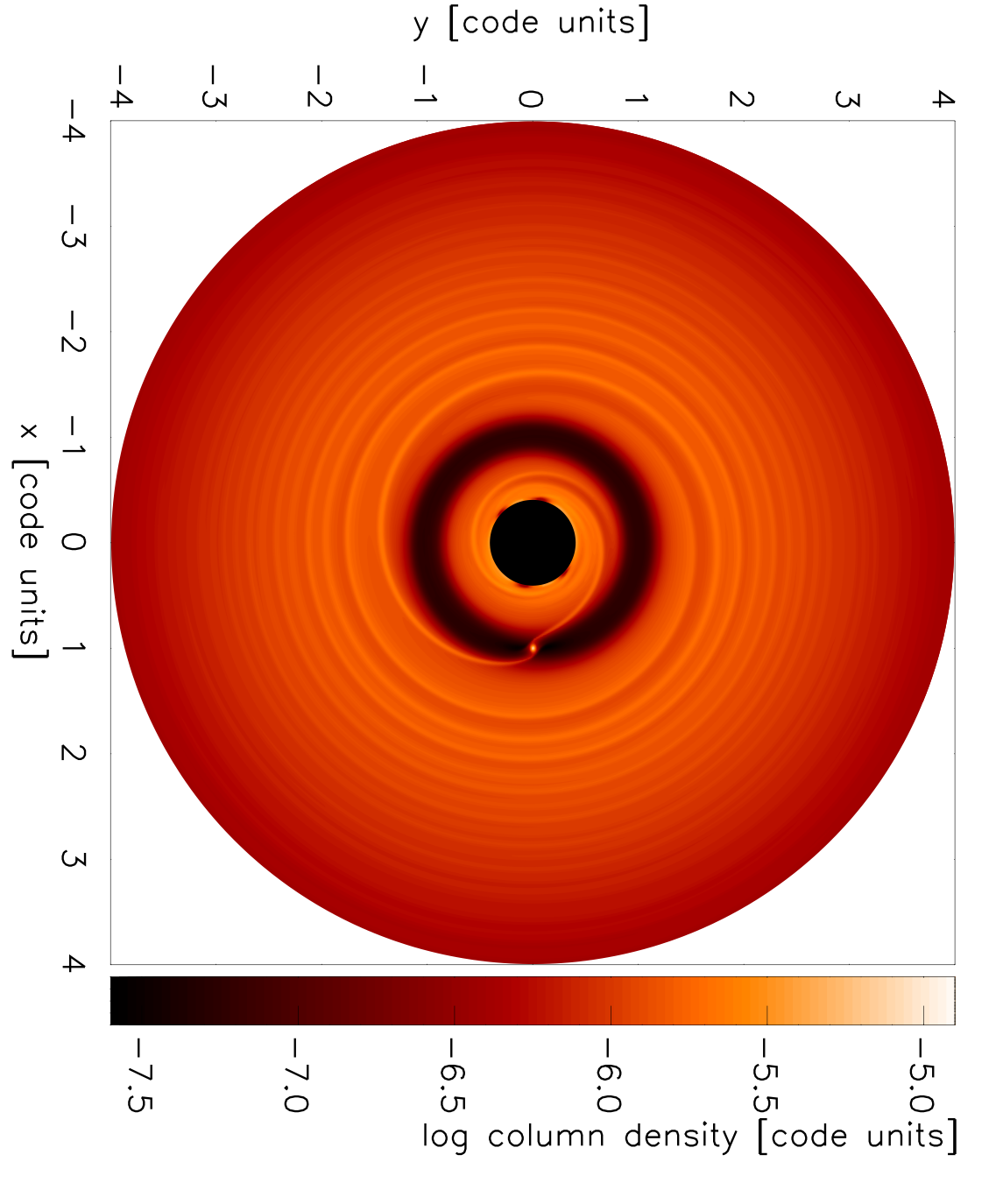}
\includegraphics[width=0.77\columnwidth]{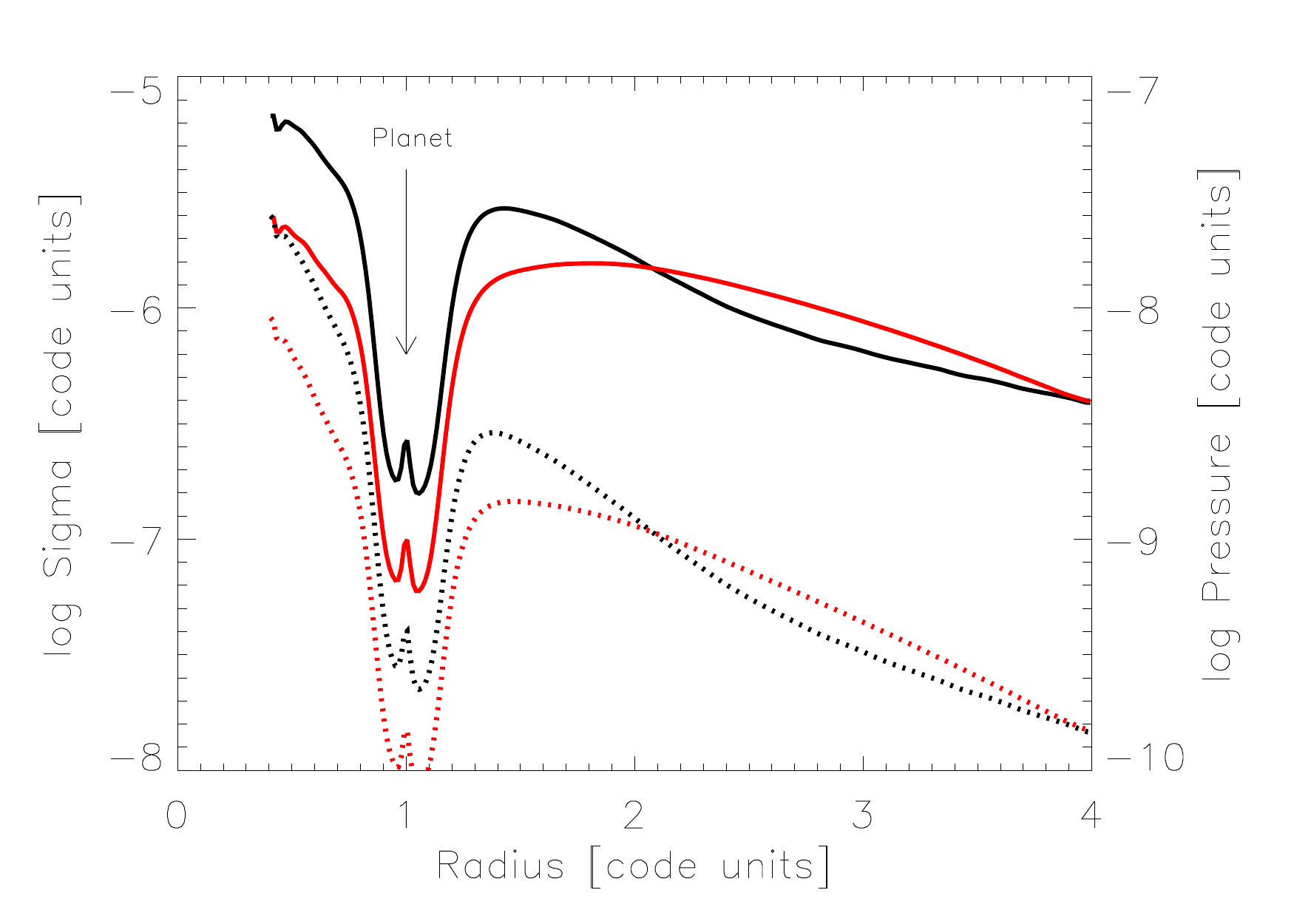}
\includegraphics[width=0.54\columnwidth,angle=90]{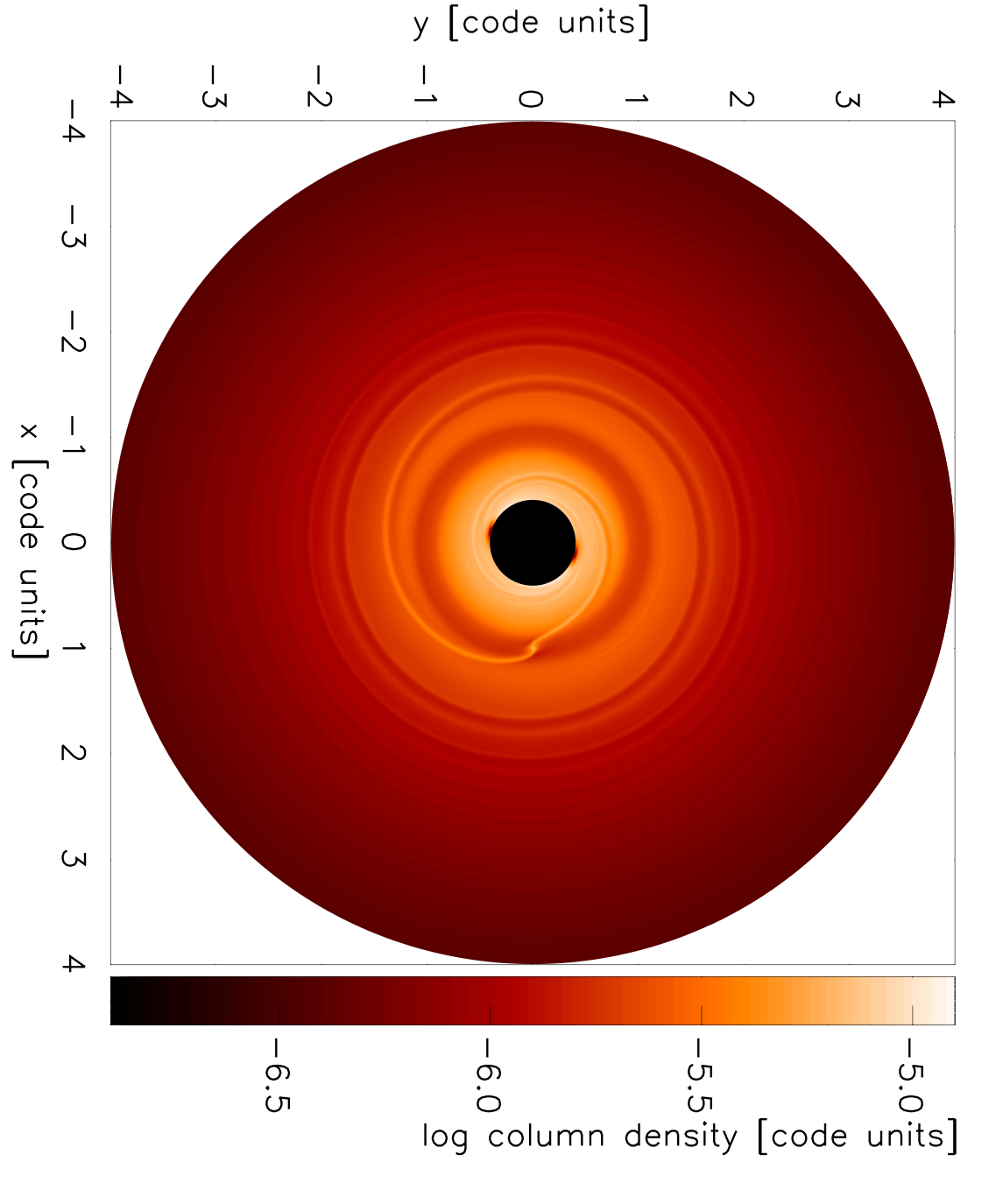}
\includegraphics[width=0.54\columnwidth,angle=90]{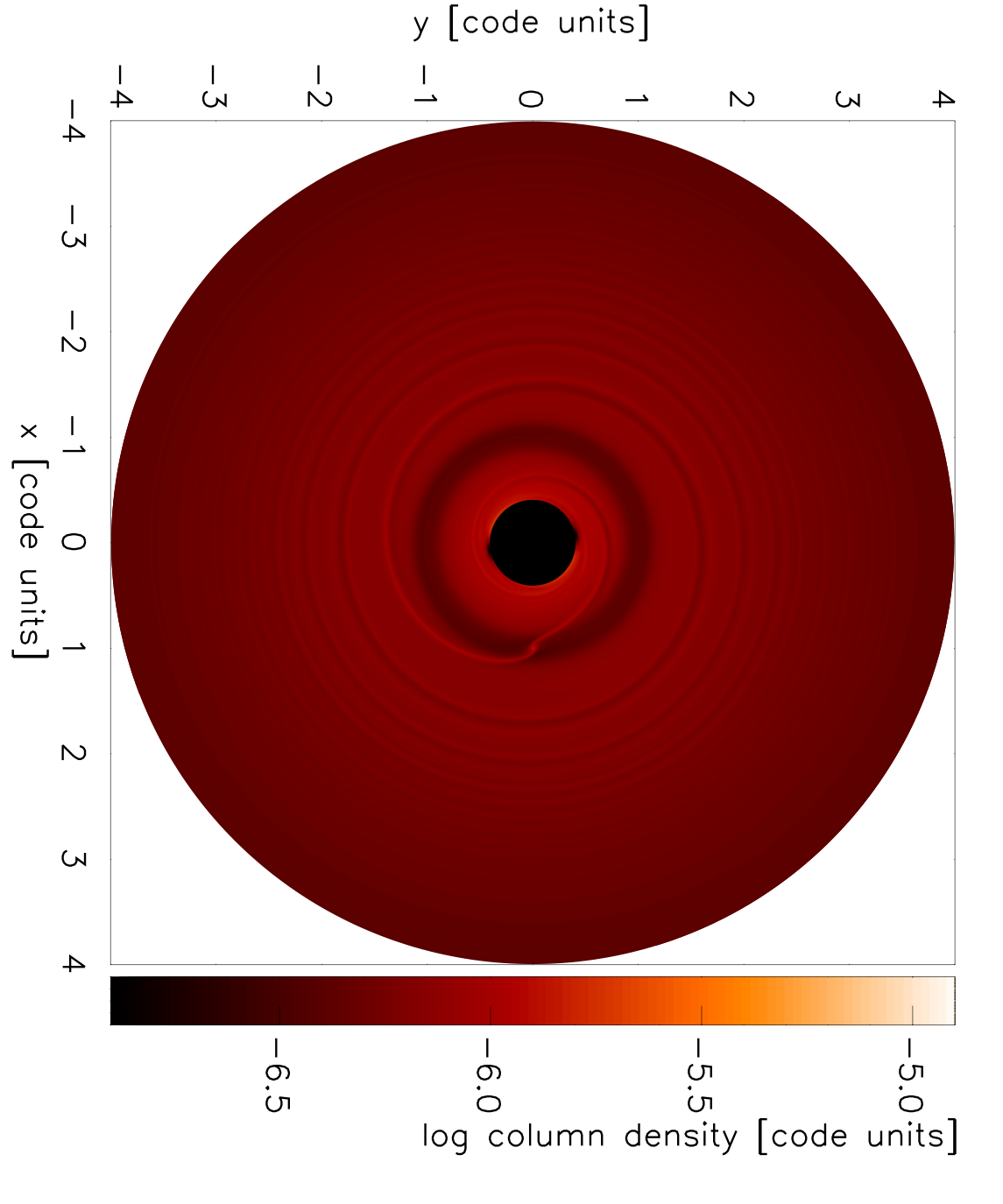}
\includegraphics[width=0.77\columnwidth]{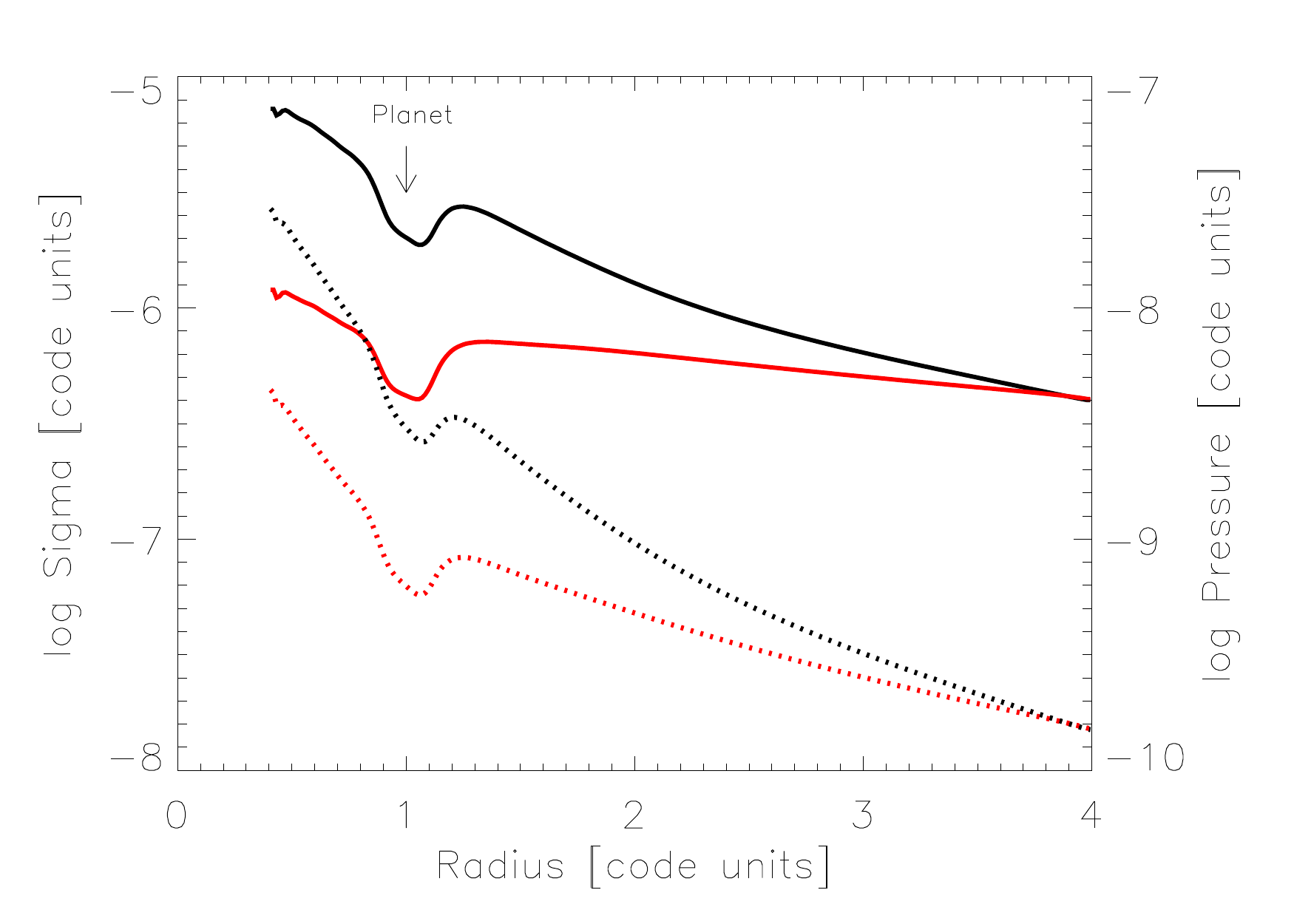}
\caption{Surface mass density rendered images of the gas disks and the azimuthally averaged surface mass density and pressure profiles of the disk with a $1 \MJup$ (top row) and $0.2 \MJup$ (bottom row) planet.  The images and profiles are at an early time (t = 500 orbits) and later time (t = 5,000 and 50,000 orbits for the $1 \MJup$ and $0.2 \MJup$ simulations, respectively).  Left column: At an earlier time the surface mass density at the outer edge of the gap is high.  Middle column: At a later time the surface density at the outer gap edge decreases.  Right column: The surface mass density (solid lines) and pressure (dotted lines) decrease between the earlier (black lines) and late times (red lines).  A pressure maximum exists at the outer gap edge where particles can be trapped and reside.  The decrease in the gas surface mass density causes an increase in the Stokes numbers of the particles and causes them to start decoupling from the gas.}
\label{fig:discs_single}
\end{figure*}

Figures~\ref{fig:discs} and~\ref{fig:discs_single} (left and middle columns) show two snapshots of the surface mass density rendered images of the disks after the planets are introduced in the multiplanet (Figure~\ref{fig:discs}) and single planet (Figure~\ref{fig:discs_single}) simulations.  The planet(s) cause either full or partial gaps to form in the high and low planet mass cases, respectively.

\subsubsection{Multiplanet simulations}
Figure~\ref{fig:discs} (left and middle columns) shows that at an earlier time a ring of gas collects in between the two planets such that its density is higher than in the co-orbital region of the planets for both planet masses that we consider.  The surface density of the gas ring then decreases due to viscous evolution, irrespective of whether the planets have opened deep or partial gaps (see also solid lines in Figure~\ref{fig:discs} right column).

Figure~\ref{fig:discs} also shows a gas surface density peak at the outermost gap edge which decreases in time, although this decrease is somewhat less significant than that in between the two planets.

\subsubsection{Single planet simulations}

Figure~\ref{fig:discs_single} (left and middle columns) shows that at an earlier time the gas surface density at the outer gap edge peaks but then decreases over time, for both planet masses considered.  Its behavior is similar to the outermost gap edge in the multiplanet case since the surface density decrease between the two times considered is somewhat less than that in the region in between two planets.

\subsection{Effects on the dust in the disk}
\label{sec:results_dust}

\begin{figure*}
\centering
  \includegraphics[width=1.0\columnwidth]{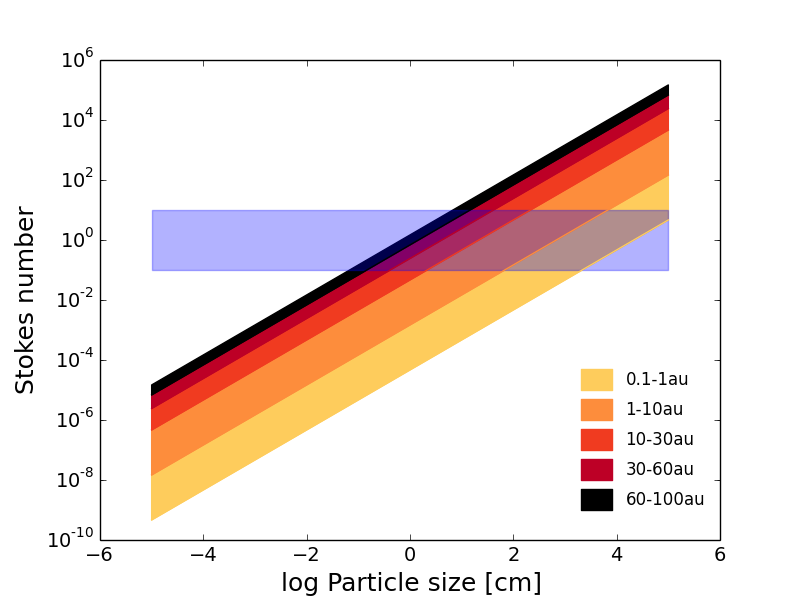}
  \includegraphics[width=1.0\columnwidth]{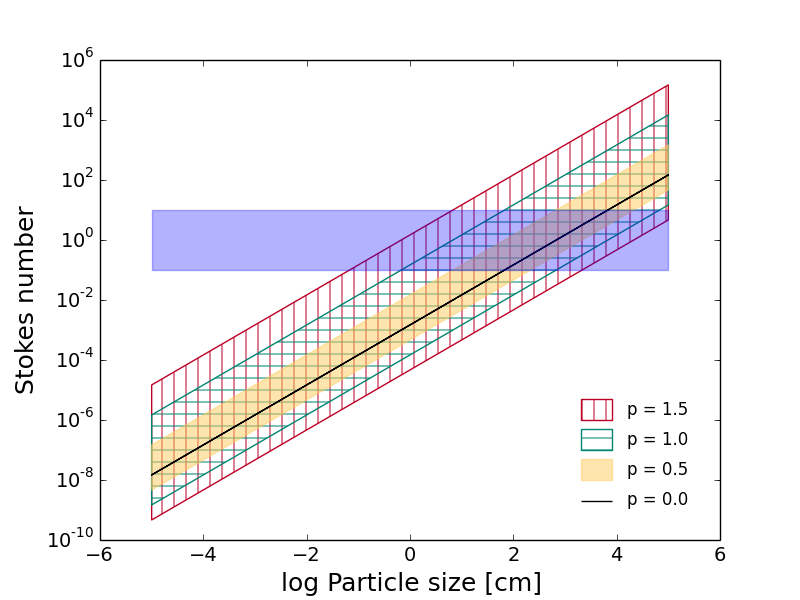}
\caption{Stokes number of particles of various sizes (diagonal bands) in disks without planets with $\Sigma (R) = 1700 (R/1AU)^{-p} \rm g/cm^2$.  The blue horizontal region highlights the particles that are marginally coupled to the gas ($\St = 0.1 - 10$).  Left panel: a MMSN disk (i.e. $p = 1.5$) divided up into various radial ranges.  In this example, the millimeter to decimeter-sized particles are marginally coupled to the gas in the outer disk and are the ones that will benefit most from being trapped in the pressure maxima.  Right panel: disks with the same normalization as the MMSN but with different surface mass density slopes.  The micron-sized particles are always coupled to the gas for disks of various surface densities.  For them to be marginally coupled the surface density would need to decrease by many orders of magnitude.}
\label{fig:Stokes}
\end{figure*}

The surface density decrease observed in the gas simulations plays an important role in the dust dynamics.  Figures~\ref{fig:discs} and~\ref{fig:discs_single} (right columns) show the azimuthally averaged surface density and pressure profiles of the gas disks at two different times.  The region in between the two planets as well as the outermost edges of the gaps in all the simulations are regions of pressure maximum where dust particles, that would ordinarily flow rapidly inwards due to radial drift, can collect.  Such particles are known to be \emph{marginally coupled} to the gas and are those that require some kind of trapping mechanism to stop them from being lost radially.  It is well known that the outer gap edge can provide such a particle trapping location.  However we show that the region in between two planets can also provide the environment for particle trapping to occur.

\subsubsection{Multiplanet case}

Figure~\ref{fig:discs} (right panel) shows that the gas surface density in between the two planets drops by almost an order of magnitude, and we expect it to continue decreasing with time until the gas disk has cleared away.  When this drop occurs, \emph{what happens to the particles that were originally trapped in the gas ring between the two planets?}  These marginally coupled particles will begin to decouple from the gas.  Consequently they will eventually be unaffected by the gas and will continue on their own Keplerian orbits, and thus remain in a ring structure.  Figure~\ref{fig:discs} (right panel) also shows that the surface density at the outermost gap edge decreases with time.  Thus, as with the region in between the two planets, the particles trapped at this location are expected to decouple, though perhaps not as easily as in the former.  Note that the smaller particles that are still coupled to the gas will follow the gas flow and so are not expected to remain in the same location.

\subsubsection{Single planet case}

Analogous to the outermost gap edge in the multiplanet case, the gas surface density at the location of the pressure maximum decreases with time, although somewhat less than in the region in between two planets.  Thus the trapping and decoupling mechanism is also expected to operate here, potentially leading to a long-lasting dust ring in the outer disk.

\subsubsection{Application to the MMSN model}

We illustrate this further using a MMSN disk model.  Figure~\ref{fig:Stokes} (left panel) shows the Stokes numbers of particles of different sizes at various disk locations.  In such a planet-free disk model the micron-sized particles are coupled to the gas ($\St \ll 1$) irrespective of radial location.  In fact, for any realistic gas-rich disk the micron-sized particles will always be coupled to the gas and will therefore follow the gas flow (see also the fact that $\St \ll 1$ for disk models with different surface mass density slopes, $p$, in Figure~\ref{fig:Stokes}, right panel).  However, millimeter and centimeter-sized particles are only marginally coupled to the gas in the outer disk and are therefore most subject to radial drift.  The locations of the pressure maxima identified above will trap these particles.  Once the gas surface mass density at these locations decreases, the Stokes numbers of these particles will increase (since $\St \propto 1/\Sigma$), and thus many will start to decouple.  We expect the millimeter- and centimeter-sized particles that had originally been marginally coupled but trapped, to decouple over time and retain their ring structure(s).  Consequently, we expect that a dust ring in the larger particles may well be a signature of this process.

\section{Discussion}
\label{sec:disc}

\begin{figure*}
\centering
  \includegraphics[width=2.0\columnwidth]{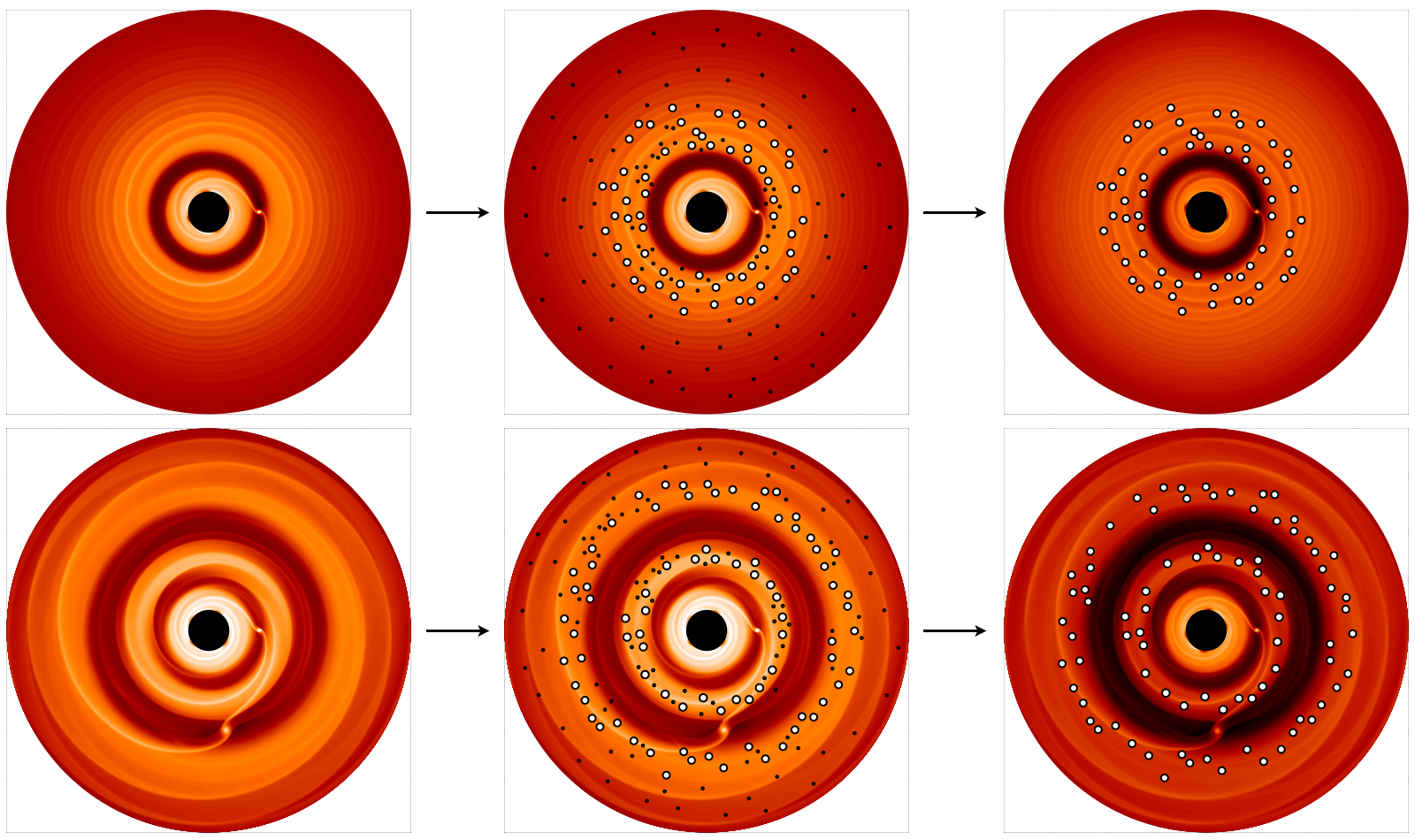}
\caption{Schematic diagram showing the stages of the proposed mechanism.  Left: A single planet (top panel) or multiple planets (bottom panel) open deep or partial gaps.  In the case of multiple planets a ring of gas forms in between the two planets.  Middle: The ring in between the two planets and the outermost gap edge are regions of pressure maxima where dust particles that would otherwise be subject to radial drift can be trapped.  Smaller grains may be trapped if the pressure maximum is strong.  Right: Decrease in gas density in the disk may cause larger trapped particles to decouple and remain in the ring structures both at the outer gap edge, and in the case of multiple planets, in the ring in between the two planets.}
\label{fig:schematic}
\end{figure*}

We propose a mechanism by which dust rings can form and last longer than gas rings.  This requires at least one body to be present in the gas-rich disk that is massive enough to at least open a partial gap.  Figure~\ref{fig:schematic} summarizes the mechanism for both the single and multiplanet cases.  This process of dust collection of marginally coupled particles in the pressure maximum followed by decoupling due to a decreased gas surface density to form a ring of dust particles is in fact scale-free and is expected to occur for different particle sizes at various radii in different disks.  While the specifics of which sizes are affected at each radii in various disk models can change, the principle mechanism is expected to operate in all disks with giant planet(s).  This may carry a stronger signature if the trapping occurs in between two planets.  Which sized particles are most subject to radial drift depends on the disk model used.  For any disk model there is always \emph{some} particle size for which this mechanism will operate.  Furthermore this general mechanism can operate with planet(s) at various radial locations provided they are massive enough to perturb the disk structure and result in a pressure maximum.  In the case of a MMSN disk around a solar mass star, the millimeter and centimeter-sized particles in the outer disk may be affected by this mechanism since they would be most subject to radial drift if a pressure maximum was not present.

As the gas disk's surface density decreases over time, the marginally coupled particles that were trapped and then decoupled will continue to become even less affected by the gas.  We expect the natural viscous evolution to aid the process of causing the rings to be long-lasting.  We also note that the pressure gradient in the disks varies in different locations.  At the outermost gap edge the gradient is steep on one side of the maximum and shallow on the other, while in between the two planets the gradient is steep on both sides of the maximum especially with higher mass planets (Figures~\ref{fig:discs} and~\ref{fig:discs_single}, right panels).  Therefore we expect the morphology of the rings in between the two planets to be different to that at the outer gap edge: in the former the rings are likely to be sharper and more confined whereas in the latter the ring may well be more extended.

The idea of a pressure maximum at an outer gap edge formed by a planet, acting to concentrate particles into it, has been considered before \citep[e.g.][]{Bryden_dust_multiplanet,Haghighipour_P_maxima,Paardekooper2004_dust,Paardekooper2006_dust,Maddison_Neptune_dust_gap,Fouchet_dustgap_easy,Fouchet_Stokes_decrease,Ayliffe_planetesimals,Pinilla_planet_dust,Zhu_dust_filtration}.  In addition disk processes associated with this e.g. particle trapping, dust filtration, excitation of Rossby Wave Instabilities, etc, may well still operate.  However the long-lasting nature of dust rings may be an alternative route if such rings are not disrupted by other disk processes.  In addition, further studies involving both gas and dust hydrodynamics are needed to establish if the ring in between two planets is disrupted by long-term disk processes.

In the single planet case it is thought that the gap edge may provide an ideal location for further planet growth \citep[e.g.][]{Bryden_dust_multiplanet}.  By extension this would also apply to the outermost gap edge in the multiplanet scenario.  With regards to the longer-term dynamics and growth of particles in the ring in between two planets, as the gas surface density decreases over time, the increased dust-to-gas ratio may result in enhanced growth or the triggering of the streaming instability.  Since the (dominant) radial drift velocity would be very small compared to that in the equivalent unperturbed disk, the overall collision velocity is expected to be small since the particles are likely to be travelling at similar speeds (unless turbulence plays a major role in increasing the collision velocities).  Therefore the collisions would be less destructive (especially if small and large particles collide since such a scenario favors growth; \citealp{Teiser_Wurm_highVcoll,Wurm_25m/s_impacts,Velocity_thresholds}).  Thus this region provides a good location for a third planet to form.  Depending on the ring size it is possible that such a planet will be lower in mass -- it is interesting to note that systems do exist where a low mass planet is sandwiched in between two higher mass planets, e.g. Kepler 20 \citep{Gautier_Kepler20}.  While this is an example of planets up to $\approx$ Neptune mass, such planets in the very inner parts of disks may open (partial) gaps due to low disk aspect ratios \citep{Baruteau_Papaloizou_Kepler} provided they have enough time to do so \citep{Malik_Meru_a}.  Furthermore, if the ring was located at large orbital distances where icy aggregates can grow more easily dust growth would be even more likely \citep{Okuzumi_RD_growth_Stokes}.  If the first two planets were in mean motion resonance then the stability of such a system whereby a third planet forms in between would need to be considered.  Conversely, if collisions within the ring were destructive this would generate smaller dust grains which may then be removed with the gas flow.  The specifics of whether growth or fragmentation dominates will depend on the specific disk and planet properties.

Previous studies of interactions between a single planet and a disk have shown gaps are easier to open in a large-sized dust population compared to the gas or small dust population \citep{Paardekooper2004_dust,Paardekooper2006_dust,Maddison_Neptune_dust_gap,Fouchet_dustgap_easy,Fouchet_Stokes_decrease,Zhu_dust_filtration}.  \cite{Paardekooper2006_dust} showed that even a Neptune mass planet could open up a gap in millimeter-sized grains.  \cite{Fouchet_dustgap_easy} and \cite{Maddison_Neptune_dust_gap} also showed this to be true for meter-sized objects.  According to the gap-opening criteria of \cite{Lin_Papaloizou_PPIII} and \cite{Crida_gap}, these are not expected to open up a gap in the gas for the disk conditions they adopted.  Therefore only a small disturbance in the gas surface density can have a significant impact on the large dust grains.  Consequently, collecting large dust particles even when only partial gaps are opened, as we suggest here, may be possible (though we note that the flow of gas and dust through the partial gaps and the region in between two planets needs to be considered in more detail via numerical simulations considering both gas and dust).

Multiplanet-disk interactions have been considered in previous studies though often in the context of forming common gaps.  Other studies do show evidence of a gas ring in between two planets during the common gap formation stage (see Figure 2 of \citealp{Kley_multiplanet}, Figure 8 of \citealp{Pierens_Nelson_2planets}, Figure 3 of \citealp{DodsonRobinson_multiplanet}).  Despite our planets being kept sufficiently far apart such that they are not expected to open a common gap, it is possible that after a long time our simulations may eventually end up with a highly diminished ring or a somewhat common gap as the disk mass decreases, or if the planet's mass was allowed to increase.  If the pressure maximum that forms in between the two planets as they open a common gap survives for long enough such that the particles most subject to radial drift can flow rapidly and become trapped (i.e. if $t_{\rm ring} > t_{\rm drift}$) then particle trapping in the ring plus decoupling may still potentially operate between the two planets even when a common gap is forming.  We note that \cite{Bryden_dust_multiplanet} also performed gas hydrodynamic simulations of two planets held at fixed circular orbits and also found the presence of a ring in between the planets.  While they considered accretion onto the planets and thus the gap widths and hence ring properties changed with time, our gas hydrodynamics simulations are in qualitative agreement in that they also found that gas is lost from the ring over time.  While they concluded that the gas rings in between the two planets are cleared in less than the lifetime of the disk the mechanism that we present simply requires particles to collect in the ring before its gas surface density decreases.  As shown later by equation~\ref{eq:tdrift_torb2}, this is of the order of tens of orbits for particles that are marginally coupled to the gas, whereas \cite{Bryden_dust_multiplanet} show the ring's mass decreases by 50\% in O(100) orbits (see their Table 2 but note that their timescales are mostly in units of the outer planet whereas ours are of the inner planet).

At the time of writing this paper, an independent study was performed by \cite{Pinilla_multiplanet} focussing on the interactions between two planets and the parent disk.  By initially performing gas hydrodynamic simulations of two planets in a disk, they take the static gas disk conditions and perform dust coagulation and fragmentation simulations.  They find that for particular values of the disk turbulence (broadly similar to the values chosen in our study) particles can collect in between the planets as well as at the outermost gap edge, in agreement with our findings.  Our results stress the importance of considering the longer-term disk evolution on the dust dynamics.  Both studies have different shortfalls: while our study does not model the coagulation and fragmentation of dust aggregates which is shown to be important in the study by these authors, \cite{Pinilla_multiplanet} do not model the time evolution of the background disk, which we show to be of importance in the long-term.  These studies together show the significance of multiplanet-disk interactions but further work is needed to understand the longer term gas and dust dynamics while also considering growth processes.

It is also worth noting that it is possible for rings to form without the presence of planets by other mechanisms \citep[see][]{Klahr_Lin_clumping_instability,Lyra_Kuchner_clumping_instability}.  However, these mechanisms require an optically thin disk.  If rings form in young protoplanetary disks (as suggested via the observations described in Section~\ref{sec:obs}) a mechanism to form them in the optically thick disk is required.

Finally, while the focus of this work has been in the context of protoplanetary disks, the mechanism is general enough to have a wide applicability in various astrophysical disk contexts where dust and gas is present and there are two large bodies in the system e.g. AGN disks.

\subsection{Further considerations}
\label{sec:further_cond}
There are some aspects to consider to ensure this mechanism is robust:\\

1) \emph{Dust excitation}
It is possible that the planet(s) cause dust to be excited so that they do not remain in the ring structures.  \cite{Hsieh_Gu_dust_excitation} estimated the timescale associated with the secular evolution of a massless particle that is perturbed by a companion as (their equation 13):

\begin{equation}
t_{\rm f} = \frac{4}{q} \frac{r}{R_{\rm p}}\left ( \frac{1}{b_{3/2}^{(2)} \left ( \alpha \right ) } \right ) \frac{1}{\Omega_{\rm k}}
\end{equation}
where $\alpha = {\rm min}(r,R_{\rm p})/{\rm max}(r,R_{\rm p})$ \citep[equation 9 of][]{Tremaine_secular}, $r$ is the location of the dust particle and $b_{3/2}^{(2)} \left ( \alpha \right )$ is the Laplace coefficient given by

\begin{equation}
b_s^{(m)} (\alpha) = \frac{2}{\pi}  \int_0^{\pi} \frac{{\rm cos}(m \phi) d \phi}{(1-2 \alpha {\rm cos} \phi + \alpha^2)^s}.
\end{equation}
The particles will be damped on the stopping timescale.  If the ratio of the stopping timescale to the timescale associated with the secular evolution of the dust $t_{\rm s}/t_{\rm f} \ll 1$ then the particles will be damped easily and the ring would not be destroyed due to the planet's interaction.  On the other hand it would be if $t_{\rm s}/t_{\rm f} \gg 1$.  We find that 

\begin{equation}
t_{\rm s}/t_{\rm f} = \St \left ( \frac{q}{4} \right ) \left ( \frac{R_{\rm p}}{r} \right ) b_{3/2}^{(2)} (\alpha).
\end{equation}
In our simulations $R_{\rm p}/r$ is of order unity: for the region in between the two planets, the planets are at $R_{\rm p} = 1$ and $R_{\rm p} = 2$ while the dust ring is located in between, and for the region at the outermost gap edge the pressure maximum is located relatively close to the planet forming it.  $b_{3/2}^{(2)}$ is between $O(1)-O(10)$ at the location of the pressure maximum both in between the two planets and in the outer disk.  Since we are interested in trapping particles that would otherwise be subject to radial drift, these particles have $\St \approx 1$.  Therefore for planetary mass objects $t_{\rm s}/t_{\rm f} \ll 1$ so the particles are not expected to be significantly excited.  It is worth noting that $t_{\rm s} \approx t_{\rm f}$ when particles are located very close to the planet (i.e. when $\alpha \longrightarrow 1$).  However, we do not expect this to be the case since the gap edges are not located extremely close to the planet (see Figures~\ref{fig:discs} and~\ref{fig:discs_single}, right panels).\\

2) \emph{Migration}
Our simulations involve planets held on fixed circular orbits while in reality planets will migrate.  Such fixed orbit simulations are more controlled and for the multiplanet case are in qualitative agreement with those in which planets migrate \emph{before} their gaps overlap (if they ever do).  We repeat the multiplanet simulations allowing the planets to migrate.  This qualitatively shows the same features i.e. the development of the gas ring in between the two planets as well as the pressure maxima between the two planets and exterior to the outermost gap, followed by the depletion of mass within these regions over time.  In particular we also note using equation~\ref{eq:ring} that two $1\MJup$ planets that are locked in a 2:1 mean motion resonance (the first first-order resonance that planets may end up in if convergent migration is not very fast; see \citealp{Baruteau_PPVI}) may migrate with a gas/dust ring in between them.\\

3) \emph{Planet and disk parameters}
We expect that the intensity, extent and lifetime of the gas/dust rings will vary with planet mass, location, disk viscosity, disk mass and surface density profile.  However, for this mechanism to operate we require the pressure maxima to form on a short enough timescale (shorter than the drift timescale) so that the particles that are most subject to radial drift can be collected easily and constrained.  The radial drift timescale of these particles, normalized by the orbital timescale at $R = 1$, can be approximated as

\begin{equation}
\frac{\tdrift}{\torb} \approx \frac{R}{u_{\rm max} \torb} ,
\label{eq:tdrift_torb}
\end{equation}
for particles that are marginally coupled to the gas, where $u_{\rm max} = \frac{1}{2 \rho \Omega} \left | \frac{dP}{dr} \right |$ is the radial drift velocity of the marginally coupled particles with $\St \approx 1$ \citep{Weidenschilling1977}.  Using the surface density and temperature structure of the disk as outlined in Table~\ref{tab:ref_disc} we find that

\begin{equation}
\frac{{\rm d}P}{{\rm d}r} = - \frac{7 c_s^2 \rho}{2 R}.
\label{eq:dP_dr}
\end{equation}
Combining equations~\ref{eq:tdrift_torb} and~\ref{eq:dP_dr} with the expression $H = c_s/\Omega$, equation~\ref{eq:tdrift_torb} becomes

\begin{equation}
\frac{\tdrift}{\torb} \approx \frac{4}{7 \pi} \left ( \frac{R}{H} \right )^2,
\label{eq:tdrift_torb2}
\end{equation}
which for our disks is $\approx 36$ orbits.  We find that the pressure maximum starts to form in less time than this, moreso for the high mass planet simulations where a deep gap opens.

As the gas surface density diminishes in time, if the pressure maxima are maintained or become stronger then the marginally coupled particles will continue to be trapped and then decouple as described in this mechanism.  In addition this may then allow enough time for further growth to occur in the ring(s).  However, if the pressure maxima weaken this may allow some particles -- especially those that are marginally coupled \emph{at that time} (which may be a different size to those that were marginally coupled and trapped when the ring(s) first formed) -- to leak out of this region, especially if the weakened pressure maxima lasts for a long time.  However, the strength of the pressure maxima and how well they are sustained in the longer term will depend on the planet and disk parameters adopted.  Furthermore, particle diffusion will also affect how much material remains in the ring structure(s) and its impact needs to be tested.\\  

4) \emph{Planet growth}
In the multiplanet case, as the planets grow in mass they will form wider gaps.  It may be possible that in such a scenario the common gap phase is more likely.  However, provided the region in between the planets forms on a short enough timescale such that it can trap particles quickly we expect the formation of the common gap will help to decouple the particles in this region quickly.

\subsection{Observational implications}
\label{sec:obs}

\subsubsection{Multiplanet case}
The proposed mechanism has clear observational predictions, which can be tested with current and/or future observing facilities.  It is known that larger particles are impacted by a pressure gradient more than the smaller particles and are thus held further away from the gas gap edge \citep[e.g.][]{Maddison_Neptune_dust_gap,Fouchet_dustgap_easy,Fouchet_Stokes_decrease,Zhu_dust_filtration}.  The gas ring in between two planets results in steep pressure gradients on both sides (moreso for the higher mass planets) and thus the large particles are held further away from the gas gap edge on \emph{both} sides of the ring.  This results in a ring of larger particles being narrower than the ring made up of smaller particles.  For spatially resolved observations this implies that sub-millimeter, millimeter and centimeter continuum observations should reveal a narrower ring with decreasing 
observing frequency. 
Scattered light observations at optical and near-infrared wavelengths, which probe mostly micron and sub-micron sized grains, 
should display a more extended ring.  However, one has to keep in mind though that these short wavelength data probe the disk surface layer and not the disk mid-plane (though note that \citealp{Zhu_2D_3D_dust_gaps} showed that dust concentrations due to gap-opening planets are very similar in 2D and 3D simulations).

Furthermore, the center of the ring in between two planets may not necessarily be in the middle-point between them.  This is because the pressure profile in the ring will not be symmetric even if  both planets have the same mass, as in our simulations: since $P \propto \Sigma\, T$, and even if the surface mass density is symmetrical 
about the ring's center, superimposing a temperature profile of $T
\propto R^{-1}$ on top of it gives an asymmetric pressure
distribution.  

In addition to the general morphology of the disk structure, the specific dust grain properties within the ring between the two planets are of particular interest from a grain growth and planet formation perspective.  Determining the spectral index ($\beta$), where $F_\nu \propto \nu^\beta$ and $F_\nu$ is the flux at a particular frequency, $\nu$, of the spatially resolved dust continuum emission from (sub-)millimeter to centimeter wavelengths, can put constraints on the typical dust grain sizes within the ring.  In addition, by comparing the emission in the ring to that in the outer disk we can study the spatial variation of $\beta$ to determine variations in the particle size distribution, and therefore establish where in the disk grain growth is more efficient in the multiplanet case.  Comparing the gas-to-dust ratio in the ring to the outer disk regions is also interesting because this will (1) affect the particle growth timescale, and (2) provide estimates on the 'age' of the ring structure.  The latter can be done by constraining the gas depletion rate in the ring due to the planets through numerical simulations involving gas and dust, and comparing the current dust-to-gas ratio value in the ring with the primordial one (in the outer disk).  

At the outermost gap edge the pressure profile is steep on one side of the maximum and shallow on the other (Figure~\ref{fig:discs}, right column) compared to the region in between two planets.  Therefore any dust rings in the former are likely to be more extended than in the latter.

All these observations require a high-spatial resolution that allows us to sufficiently resolve the individual disk regions. 
Current (sub-)mm facilities (SMA, CARMA, PdBI) can provide angular resolutions of at best $\approx$0.2$\arcsec$, while ALMA Cycle 2 and Jansky VLA (at 15\,GHz) provide 0.13$\arcsec$ in their most extended configurations.  However, ALMA's full capabilities and SKA1 (at 16\,GHz) will provide angular resolutions of 0.04$\arcsec$, which translates into structures of 6\,au in size at the distance of the nearest star-forming regions ($\sim 150$\,pc). This resolution is comparable to that of SPHERE/ZIMPOL, the optical imaging polarimeter for the VLT \citep{schmid2006}, which will trace the disk surface layer in scattered light. All these facilities are in the southern hemisphere making a multi-wavelength comparison of the same sources possible. 

HD100546 and HD169142 may be ideal testbeds to determine whether the proposed mechanism works.  For HD100546, where the dust ring in between the two protoplanetary candidates appears to extend from $\approx 14-60$\,au \citep{Pineda_HD100546,Walsh_HD100546}, ALMA Cycle 2 already provides a high enough  
spatial resolution (0.12$\arcsec$ at 345 and 460\,GHz) to 
test whether the apparent width of the ring changes as a function of observing wavelength.  These observations would simultaneously probe the dust (continuum) and gas (CO line), and therefore such data would allow us to put constraints on the gas content 
and dust grain properties within the ring compared to the outer disk (e.g. gas and dust mass, as well as dust sizes and compositions via the spectral index).  Furthermore, while direct images strongly support the existence of an outer planet \citep{HD100546b,Currie_HD100546b_confirm,Quanz_HD100546_confirmation}, the inner planet has thus far only been inferred indirectly \citep[e.g.,][]{Brittain_HD100546_candidate,Brittain_HD100546_candidate2}. New high-contrast imaging instruments such as SPHERE at the VLT \citep{Beuzit_SPHERE} or GPI at Gemini Observatory \citep{Macintosh_GPI} may provide sufficient contrast between 0.1$\arcsec$-0.15$\arcsec$ to directly detect the object and provide better constraints on the planet properties which will then inform models of the observed ring structures. 

The ring in HD169142 is neither resolved in scattered light observations
with a resolution of $\sim$0.1$\arcsec$ \citep{Quanz_doughnut}, nor in the 7-mm EVLA observation by
\cite{Osorio_HD169142_7mm}. It will therefore be challenging to determine the
ring width as a function of wavelength in the (sub)-mm
regime. However, further characterization of the inner protoplanet
candidate \citep{Reggiani_HD169142,Biller_HD169142} through
emission-lines (e.g., $\rm H\alpha$) and multi-color photometry will
help separate the flux coming from the central object and a possible
circumplanetary disk, and may ultimately shed light on the ring properties. Furthermore, ALMA could test whether the
apparent clumpiness of the ring emission seen in the EVLA data
is real and also if the suspected emission source in the disk gap at
$\approx 50$\,au can be confirmed.  The former point would hint towards ongoing
grain processing within the dust ring and the latter point would
support the existence of an outer companion.  Finally, as with HD100546, SPHERE and GPI may constrain the properties of the planet detected in the inner cavity of HD169142, and hence inform us about the ring properties.

\subsubsection{Single planet case}

To date, rings have only been detected in disks which are thought to harbor multiple planets (aside from HL Tau whose planet nature is currently not known).  However, spatially resolved images of disks is still a new and developing area.  If rings in disks with single planets were to exist, the morphology is expected to be similar to those in the outer part of a disk that harbors multiple planets.  Observed disks with structure, which so far have not had any planets detected, have shown varying gap sizes depending on the observing wavelengths \citep[e.g.][]{Garufi_SAO206462} amongst other structures such as strong azimuthal brightness asymmetries \citep[e.g.][]{Nienke_dust_trap,Perez_TD_asymmetries} and spiral structures \citep[e.g.][]{Boccaletti_spirals_HD100546,Avenhaus_HD142527}.  Such phenomena may be due to the interaction between a disk and an orbiting planet.  The dust-gas dynamics involved in the former, i.e. that the dust is trapped at the pressure maximum \citep[e.g.][]{Zhu_dust_filtration,Pinilla_pressure_bump,deJuanOvelar_TDs}, also applies to our mechanism.

These examples are of disks where a planet has so far not being detected.  The LkCa15 system is the only system (aside from the above-mentioned HD169142 and HD100546), where a young planet candidate has directly been detected. This object orbits at $\sim$11 au within the gap of the transition disk surrounding the star \citep{LkCa15_protoplanet} and the gap has been resolved in scattered light and at sub-mm wavelength \citep{Thalmann_LkCa15_ecc,Andrews_LkCa15}. The current data do not have sufficient spatial resolution to reveal significantly different cavity sizes or the potential for ring-like structures.

\subsubsection{Long-lasting nature of dust rings}

Since our results suggest that dust rings should be longer-lasting than gas rings, unless some mechanism acts to remove or re-distribute these particles, they may remain at least until the phase when the disk transitions from the gas-rich protoplanetary stage to the gas-poor debris disk stage.  Future observations of gas and dust in the millimeter in young debris disks, or equivalently older protoplanetary disks, may determine whether rings are commonly present in such disks that are in transition.  Furthermore, to provide a statistical view on how common and long-lasting ring structures are, many sources would need to be observed at high resolution.  Combined with any potential planet detections in such disks, this may provide clues on the long-lasting nature of dust rings formed due to planet-disk interactions.

The recent image of the rings in the young HL Tau disk ($\lesssim 10^6$ years) also hints at the long-lasting nature of dust rings, albeit from the young disk perspective.  If such ring structures are due to planet-disk interactions, this must take place at a young stage when the planets may still be forming, and requires a mechanism that can concentrate the dust into rings rapidly. Depending on the exact location of the rings and using an upper age of $10^6$ years and a mass of $0.5 \Msolar$ for HL Tau, the rings would need to form within a timescale of $\rm O(10^2)-O(10^4)$ orbits at radii between 10-200 au.  If the disk age is an order of magnitude smaller, these timescales would reduce by a factor of 10.  Our results suggest that ring formation within this time is certainly feasible if planets are present in the disk.

\section{Conclusions}
\label{sec:conc}

We perform 2D hydrodynamical simulations and combine this with analytics to investigate the gas and dust ring structures in a gas-rich protoplanetary disk when a single or two planets are present.  In the multiplanet case the planets are placed sufficiently far apart such that the gaps (partial or deep) do not overlap.  Through the hydrodynamical simulations we find that a ring of gas exists in between the two planets with a higher surface density than either side of the ring.  This leads to a pressure maximum whose prominence increases with planet mass.  In addition, pressure maxima form at the outermost gap edge in both the single and multiplanet simulations.  These pressure maxima are where dust grains, especially those that are most subject to radial drift, are likely to collect and become trapped in the ring structure(s).  Over time the gas density decreases while still maintaining the pressure maxima.  Therefore the larger particles are expected to start decoupling from the gas.  Consequently, as the gas ring disappears over time, the dust would be less affected and may remain in the ring structure(s).  The gas density decrease appears to be more prominent in the region in between two planets, so this trapping and decoupling process may be more important in the multiplanet case.  We therefore propose that dust rings structures should last longer than gas ring structures.  Furthermore, the ring features seen in recently observed gas-rich disks (HD100546, HD169142 and HL Tau) may well be signposts of multiple planet-disk interactions.

\acknowledgments
We thank Doug Lin, Amaury Triaud, Lucio Mayer, Ali Rahmati and Olja Pani\'c for inspiring discussions.  The calculations reported here were performed using the {\sc brutus} cluster at ETH Z\"urich.  FM was supported by the ETH Zurich Postdoctoral Fellowship Programme as well as by the Marie Curie Actions for People COFUND program.  This work has been supported by the DISCSIM project, grant agreement 341137 funded by the European Research Council under ERC-2013-ADG.  SQ and MR acknowledge financial support from the Swiss National Science Foundation in the frame of the National Centre for Competence in Research 'PlanetS'.  JEP is supported by the Swiss National Science Foundation, project number CRSII2\_141880.


\begin{thebibliography}{60}
\expandafter\ifx\csname natexlab\endcsname\relax\def\natexlab#1{#1}\fi
\expandafter\ifx\csname bibnamefont\endcsname\relax
  \def\bibnamefont#1{#1}\fi
\expandafter\ifx\csname bibfnamefont\endcsname\relax
  \def\bibfnamefont#1{#1}\fi
\expandafter\ifx\csname citenamefont\endcsname\relax
  \def\citenamefont#1{#1}\fi
\expandafter\ifx\csname url\endcsname\relax
  \def\url#1{\texttt{#1}}\fi
\expandafter\ifx\csname urlprefix\endcsname\relax\def\urlprefix{URL }\fi
\providecommand{\bibinfo}[2]{#2}
\providecommand{\eprint}[2][]{\url{#2}}

\bibitem[{\citenamefont{{Andrews} et~al.}(2011)\citenamefont{{Andrews},
  {Rosenfeld}, {Wilner}, and {Bremer}}}]{Andrews_LkCa15}
\bibinfo{author}{\bibfnamefont{S.~M.} \bibnamefont{{Andrews}}},
  \bibinfo{author}{\bibfnamefont{K.~A.} \bibnamefont{{Rosenfeld}}},
  \bibinfo{author}{\bibfnamefont{D.~J.} \bibnamefont{{Wilner}}},
  \bibnamefont{and} \bibinfo{author}{\bibfnamefont{M.}~\bibnamefont{{Bremer}}},
  \bibinfo{journal}{\apjl} \textbf{\bibinfo{volume}{742}}, \bibinfo{eid}{L5}
  (\bibinfo{year}{2011}), \eprint{1110.3865}.

\bibitem[{\citenamefont{{Avenhaus}
  et~al.}(2014{\natexlab{a}})\citenamefont{{Avenhaus}, {Quanz}, {Schmid},
  {Meyer}, {Garufi}, {Wolf}, and {Dominik}}}]{Avenhaus_HD142527}
\bibinfo{author}{\bibfnamefont{H.}~\bibnamefont{{Avenhaus}}},
  \bibinfo{author}{\bibfnamefont{S.~P.} \bibnamefont{{Quanz}}},
  \bibinfo{author}{\bibfnamefont{H.~M.} \bibnamefont{{Schmid}}},
  \bibinfo{author}{\bibfnamefont{M.~R.} \bibnamefont{{Meyer}}},
  \bibinfo{author}{\bibfnamefont{A.}~\bibnamefont{{Garufi}}},
  \bibinfo{author}{\bibfnamefont{S.}~\bibnamefont{{Wolf}}}, \bibnamefont{and}
  \bibinfo{author}{\bibfnamefont{C.}~\bibnamefont{{Dominik}}},
  \bibinfo{journal}{\apj} \textbf{\bibinfo{volume}{781}}, \bibinfo{eid}{87}
  (\bibinfo{year}{2014}{\natexlab{a}}), \eprint{1311.7088}.

\bibitem[{\citenamefont{{Avenhaus}
  et~al.}(2014{\natexlab{b}})\citenamefont{{Avenhaus}, {Quanz}, {Meyer},
  {Brittain}, {Carr}, and {Najita}}}]{Avenhaus_HD100546_cavity}
\bibinfo{author}{\bibfnamefont{H.}~\bibnamefont{{Avenhaus}}},
  \bibinfo{author}{\bibfnamefont{S.~P.} \bibnamefont{{Quanz}}},
  \bibinfo{author}{\bibfnamefont{M.~R.} \bibnamefont{{Meyer}}},
  \bibinfo{author}{\bibfnamefont{S.~D.} \bibnamefont{{Brittain}}},
  \bibinfo{author}{\bibfnamefont{J.~S.} \bibnamefont{{Carr}}},
  \bibnamefont{and} \bibinfo{author}{\bibfnamefont{J.~R.}
  \bibnamefont{{Najita}}}, \bibinfo{journal}{\apj}
  \textbf{\bibinfo{volume}{790}}, \bibinfo{eid}{56}
  (\bibinfo{year}{2014}{\natexlab{b}}), \eprint{1405.6120}.

\bibitem[{\citenamefont{{Ayliffe} et~al.}(2012)\citenamefont{{Ayliffe},
  {Laibe}, {Price}, and {Bate}}}]{Ayliffe_planetesimals}
\bibinfo{author}{\bibfnamefont{B.~A.} \bibnamefont{{Ayliffe}}},
  \bibinfo{author}{\bibfnamefont{G.}~\bibnamefont{{Laibe}}},
  \bibinfo{author}{\bibfnamefont{D.~J.} \bibnamefont{{Price}}},
  \bibnamefont{and} \bibinfo{author}{\bibfnamefont{M.~R.}
  \bibnamefont{{Bate}}}, \bibinfo{journal}{\mnras}
  \textbf{\bibinfo{volume}{423}}, \bibinfo{pages}{1450} (\bibinfo{year}{2012}),
  \eprint{1203.4953}.




\bibitem[{\citenamefont{{Baruteau} and
  {Papaloizou}}(2013)}]{Baruteau_Papaloizou_Kepler}
\bibinfo{author}{\bibfnamefont{C.}~\bibnamefont{{Baruteau}}} \bibnamefont{and}
  \bibinfo{author}{\bibfnamefont{J.~C.~B.} \bibnamefont{{Papaloizou}}},
  \bibinfo{journal}{\apj} \textbf{\bibinfo{volume}{778}}, \bibinfo{eid}{7}
  (\bibinfo{year}{2013}), \eprint{1301.0779}.

\bibitem[{\citenamefont{{Baruteau} et~al.}(2013)\citenamefont{{Baruteau},
  {Crida}, {Paardekooper}, {Masset}, {Guilet}, {Bitsch}, {Nelson}, {Kley}, and
  {Papaloizou}}}]{Baruteau_PPVI}
\bibinfo{author}{\bibfnamefont{C.}~\bibnamefont{{Baruteau}}},
  \bibinfo{author}{\bibfnamefont{A.}~\bibnamefont{{Crida}}},
  \bibinfo{author}{\bibfnamefont{S.-J.} \bibnamefont{{Paardekooper}}},
  \bibinfo{author}{\bibfnamefont{F.}~\bibnamefont{{Masset}}},
  \bibinfo{author}{\bibfnamefont{J.}~\bibnamefont{{Guilet}}},
  \bibinfo{author}{\bibfnamefont{B.}~\bibnamefont{{Bitsch}}},
  \bibinfo{author}{\bibfnamefont{R.~P.} \bibnamefont{{Nelson}}},
  \bibinfo{author}{\bibfnamefont{W.}~\bibnamefont{{Kley}}}, \bibnamefont{and}
  \bibinfo{author}{\bibfnamefont{J.~C.~B.} \bibnamefont{{Papaloizou}}},
  \bibinfo{journal}{ArXiv e-prints}  (\bibinfo{year}{2013}),
  \eprint{1312.4293}.

\bibitem[{\citenamefont{{Beuzit} et~al.}(2006)\citenamefont{{Beuzit}, {Feldt},
  {Dohlen}, {Mouillet}, {Puget}, {Antichi}, {Baruffolo}, {Baudoz}, {Berton},
  {Boccaletti} et~al.}}]{Beuzit_SPHERE}
\bibinfo{author}{\bibfnamefont{J.-L.} \bibnamefont{{Beuzit}}},
  \bibinfo{author}{\bibfnamefont{M.}~\bibnamefont{{Feldt}}},
  \bibinfo{author}{\bibfnamefont{K.}~\bibnamefont{{Dohlen}}},
  \bibinfo{author}{\bibfnamefont{D.}~\bibnamefont{{Mouillet}}},
  \bibinfo{author}{\bibfnamefont{P.}~\bibnamefont{{Puget}}},
  \bibinfo{author}{\bibfnamefont{J.}~\bibnamefont{{Antichi}}},
  \bibinfo{author}{\bibfnamefont{A.}~\bibnamefont{{Baruffolo}}},
  \bibinfo{author}{\bibfnamefont{P.}~\bibnamefont{{Baudoz}}},
  \bibinfo{author}{\bibfnamefont{A.}~\bibnamefont{{Berton}}},
  \bibinfo{author}{\bibfnamefont{A.}~\bibnamefont{{Boccaletti}}},
  \bibnamefont{et~al.}, \bibinfo{journal}{The Messenger}
  \textbf{\bibinfo{volume}{125}}, \bibinfo{pages}{29} (\bibinfo{year}{2006}).

\bibitem[{\citenamefont{{Biller} et~al.}(2014)\citenamefont{{Biller}, {Males},
  {Rodigas}, {Morzinski}, {Close}, {Juh{\'a}sz}, {Follette}, {Lacour},
  {Benisty}, {Sicilia-Aguilar} et~al.}}]{Biller_HD169142}
\bibinfo{author}{\bibfnamefont{B.~A.} \bibnamefont{{Biller}}},
  \bibinfo{author}{\bibfnamefont{J.}~\bibnamefont{{Males}}},
  \bibinfo{author}{\bibfnamefont{T.}~\bibnamefont{{Rodigas}}},
  \bibinfo{author}{\bibfnamefont{K.}~\bibnamefont{{Morzinski}}},
  \bibinfo{author}{\bibfnamefont{L.~M.} \bibnamefont{{Close}}},
  \bibinfo{author}{\bibfnamefont{A.}~\bibnamefont{{Juh{\'a}sz}}},
  \bibinfo{author}{\bibfnamefont{K.~B.} \bibnamefont{{Follette}}},
  \bibinfo{author}{\bibfnamefont{S.}~\bibnamefont{{Lacour}}},
  \bibinfo{author}{\bibfnamefont{M.}~\bibnamefont{{Benisty}}},
  \bibinfo{author}{\bibfnamefont{A.}~\bibnamefont{{Sicilia-Aguilar}}},
  \bibnamefont{et~al.}, \bibinfo{journal}{ArXiv e-prints}
  (\bibinfo{year}{2014}), \eprint{1408.0794}.

\bibitem[{\citenamefont{{Boccaletti} et~al.}(2013)\citenamefont{{Boccaletti},
  {Pantin}, {Lagrange}, {Augereau}, {Meheut}, and
  {Quanz}}}]{Boccaletti_spirals_HD100546}
\bibinfo{author}{\bibfnamefont{A.}~\bibnamefont{{Boccaletti}}},
  \bibinfo{author}{\bibfnamefont{E.}~\bibnamefont{{Pantin}}},
  \bibinfo{author}{\bibfnamefont{A.-M.} \bibnamefont{{Lagrange}}},
  \bibinfo{author}{\bibfnamefont{J.-C.} \bibnamefont{{Augereau}}},
  \bibinfo{author}{\bibfnamefont{H.}~\bibnamefont{{Meheut}}}, \bibnamefont{and}
  \bibinfo{author}{\bibfnamefont{S.~P.} \bibnamefont{{Quanz}}},
  \bibinfo{journal}{\aap} \textbf{\bibinfo{volume}{560}}, \bibinfo{eid}{A20}
  (\bibinfo{year}{2013}), \eprint{1310.7092}.

\bibitem[{\citenamefont{{Brittain} et~al.}(2013)\citenamefont{{Brittain},
  {Najita}, {Carr}, {Liskowsky}, {Troutman}, and
  {Doppmann}}}]{Brittain_HD100546_candidate}
\bibinfo{author}{\bibfnamefont{S.~D.} \bibnamefont{{Brittain}}},
  \bibinfo{author}{\bibfnamefont{J.~R.} \bibnamefont{{Najita}}},
  \bibinfo{author}{\bibfnamefont{J.~S.} \bibnamefont{{Carr}}},
  \bibinfo{author}{\bibfnamefont{J.}~\bibnamefont{{Liskowsky}}},
  \bibinfo{author}{\bibfnamefont{M.~R.} \bibnamefont{{Troutman}}},
  \bibnamefont{and} \bibinfo{author}{\bibfnamefont{G.~W.}
  \bibnamefont{{Doppmann}}}, \bibinfo{journal}{\apj}
  \textbf{\bibinfo{volume}{767}}, \bibinfo{eid}{159} (\bibinfo{year}{2013}),
  \eprint{1311.5647}.

\bibitem[{\citenamefont{{Brittain} et~al.}(2014)\citenamefont{{Brittain},
  {Carr}, {Najita}, {Quanz}, and {Meyer}}}]{Brittain_HD100546_candidate2}
\bibinfo{author}{\bibfnamefont{S.~D.} \bibnamefont{{Brittain}}},
  \bibinfo{author}{\bibfnamefont{J.~S.} \bibnamefont{{Carr}}},
  \bibinfo{author}{\bibfnamefont{J.~R.} \bibnamefont{{Najita}}},
  \bibinfo{author}{\bibfnamefont{S.~P.} \bibnamefont{{Quanz}}},
  \bibnamefont{and} \bibinfo{author}{\bibfnamefont{M.~R.}
  \bibnamefont{{Meyer}}}, \bibinfo{journal}{\apj}
  \textbf{\bibinfo{volume}{791}}, \bibinfo{eid}{136} (\bibinfo{year}{2014}),
  \eprint{1409.0804}.

\bibitem[{\citenamefont{{Bryden} et~al.}(2000)\citenamefont{{Bryden},
  {R{\'o}{\.z}yczka}, {Lin}, and {Bodenheimer}}}]{Bryden_dust_multiplanet}
\bibinfo{author}{\bibfnamefont{G.}~\bibnamefont{{Bryden}}},
  \bibinfo{author}{\bibfnamefont{M.}~\bibnamefont{{R{\'o}{\.z}yczka}}},
  \bibinfo{author}{\bibfnamefont{D.~N.~C.} \bibnamefont{{Lin}}},
  \bibnamefont{and}
  \bibinfo{author}{\bibfnamefont{P.}~\bibnamefont{{Bodenheimer}}},
  \bibinfo{journal}{\apj} \textbf{\bibinfo{volume}{540}}, \bibinfo{pages}{1091}
  (\bibinfo{year}{2000}).

\bibitem[{\citenamefont{{Crida} et~al.}(2006)\citenamefont{{Crida},
  {Morbidelli}, and {Masset}}}]{Crida_gap}
\bibinfo{author}{\bibfnamefont{A.}~\bibnamefont{{Crida}}},
  \bibinfo{author}{\bibfnamefont{A.}~\bibnamefont{{Morbidelli}}},
  \bibnamefont{and} \bibinfo{author}{\bibfnamefont{F.}~\bibnamefont{{Masset}}},
  \bibinfo{journal}{\icarus} \textbf{\bibinfo{volume}{181}},
  \bibinfo{pages}{587} (\bibinfo{year}{2006}), \eprint{astro-ph/0511082}.

\bibitem[{\citenamefont{{Currie} et~al.}(2014)\citenamefont{{Currie}, {Muto},
  {Kudo}, {Honda}, {Brandt}, {Grady}, {Fukagawa}, {Burrows}, {Janson},
  {Kuzuhara} et~al.}}]{Currie_HD100546b_confirm}
\bibinfo{author}{\bibfnamefont{T.}~\bibnamefont{{Currie}}},
  \bibinfo{author}{\bibfnamefont{T.}~\bibnamefont{{Muto}}},
  \bibinfo{author}{\bibfnamefont{T.}~\bibnamefont{{Kudo}}},
  \bibinfo{author}{\bibfnamefont{M.}~\bibnamefont{{Honda}}},
  \bibinfo{author}{\bibfnamefont{T.~D.} \bibnamefont{{Brandt}}},
  \bibinfo{author}{\bibfnamefont{C.}~\bibnamefont{{Grady}}},
  \bibinfo{author}{\bibfnamefont{M.}~\bibnamefont{{Fukagawa}}},
  \bibinfo{author}{\bibfnamefont{A.}~\bibnamefont{{Burrows}}},
  \bibinfo{author}{\bibfnamefont{M.}~\bibnamefont{{Janson}}},
  \bibinfo{author}{\bibfnamefont{M.}~\bibnamefont{{Kuzuhara}}},
  \bibnamefont{et~al.}, \bibinfo{journal}{ArXiv e-prints}
  (\bibinfo{year}{2014}), \eprint{1411.0315}.

\bibitem[{\citenamefont{{de Juan Ovelar} et~al.}(2013)\citenamefont{{de Juan
  Ovelar}, {Min}, {Dominik}, {Thalmann}, {Pinilla}, {Benisty}, and
  {Birnstiel}}}]{deJuanOvelar_TDs}
\bibinfo{author}{\bibfnamefont{M.}~\bibnamefont{{de Juan Ovelar}}},
  \bibinfo{author}{\bibfnamefont{M.}~\bibnamefont{{Min}}},
  \bibinfo{author}{\bibfnamefont{C.}~\bibnamefont{{Dominik}}},
  \bibinfo{author}{\bibfnamefont{C.}~\bibnamefont{{Thalmann}}},
  \bibinfo{author}{\bibfnamefont{P.}~\bibnamefont{{Pinilla}}},
  \bibinfo{author}{\bibfnamefont{M.}~\bibnamefont{{Benisty}}},
  \bibnamefont{and}
  \bibinfo{author}{\bibfnamefont{T.}~\bibnamefont{{Birnstiel}}},
  \bibinfo{journal}{\aap} \textbf{\bibinfo{volume}{560}}, \bibinfo{eid}{A111}
  (\bibinfo{year}{2013}), \eprint{1309.1039}.

\bibitem[{\citenamefont{{Dodson-Robinson} and
  {Salyk}}(2011)}]{DodsonRobinson_multiplanet}
\bibinfo{author}{\bibfnamefont{S.~E.} \bibnamefont{{Dodson-Robinson}}}
  \bibnamefont{and} \bibinfo{author}{\bibfnamefont{C.}~\bibnamefont{{Salyk}}},
  \bibinfo{journal}{\apj} \textbf{\bibinfo{volume}{738}}, \bibinfo{eid}{131}
  (\bibinfo{year}{2011}), \eprint{1106.4824}.


\bibitem[{\citenamefont{{Fouchet} et~al.}(2007)\citenamefont{{Fouchet},
  {Maddison}, {Gonzalez}, and {Murray}}}]{Fouchet_dustgap_easy}
\bibinfo{author}{\bibfnamefont{L.}~\bibnamefont{{Fouchet}}},
  \bibinfo{author}{\bibfnamefont{S.~T.} \bibnamefont{{Maddison}}},
  \bibinfo{author}{\bibfnamefont{J.-F.} \bibnamefont{{Gonzalez}}},
  \bibnamefont{and} \bibinfo{author}{\bibfnamefont{J.~R.}
  \bibnamefont{{Murray}}}, \bibinfo{journal}{\aap}
  \textbf{\bibinfo{volume}{474}}, \bibinfo{pages}{1037} (\bibinfo{year}{2007}),
  \eprint{0708.4110}.

\bibitem[{\citenamefont{{Fouchet} et~al.}(2010)\citenamefont{{Fouchet},
  {Gonzalez}, and {Maddison}}}]{Fouchet_Stokes_decrease}
\bibinfo{author}{\bibfnamefont{L.}~\bibnamefont{{Fouchet}}},
  \bibinfo{author}{\bibfnamefont{J.-F.} \bibnamefont{{Gonzalez}}},
  \bibnamefont{and} \bibinfo{author}{\bibfnamefont{S.~T.}
  \bibnamefont{{Maddison}}}, \bibinfo{journal}{\aap}
  \textbf{\bibinfo{volume}{518}}, \bibinfo{eid}{A16} (\bibinfo{year}{2010}),
  \eprint{1005.4557}.

\bibitem[{\citenamefont{{Garaud} et~al.}(2013)\citenamefont{{Garaud}, {Meru},
  {Galvagni}, and {Olczak}}}]{Garaud_vel_pdf}
\bibinfo{author}{\bibfnamefont{P.}~\bibnamefont{{Garaud}}},
  \bibinfo{author}{\bibfnamefont{F.}~\bibnamefont{{Meru}}},
  \bibinfo{author}{\bibfnamefont{M.}~\bibnamefont{{Galvagni}}},
  \bibnamefont{and} \bibinfo{author}{\bibfnamefont{C.}~\bibnamefont{{Olczak}}},
  \bibinfo{journal}{\apj} \textbf{\bibinfo{volume}{764}}, \bibinfo{eid}{146}
  (\bibinfo{year}{2013}), \eprint{1209.0013}.

\bibitem[{\citenamefont{{Garufi} et~al.}(2013)\citenamefont{{Garufi}, {Quanz},
  {Avenhaus}, {Buenzli}, {Dominik}, {Meru}, {Meyer}, {Pinilla}, {Schmid}, and
  {Wolf}}}]{Garufi_SAO206462}
\bibinfo{author}{\bibfnamefont{A.}~\bibnamefont{{Garufi}}},
  \bibinfo{author}{\bibfnamefont{S.~P.} \bibnamefont{{Quanz}}},
  \bibinfo{author}{\bibfnamefont{H.}~\bibnamefont{{Avenhaus}}},
  \bibinfo{author}{\bibfnamefont{E.}~\bibnamefont{{Buenzli}}},
  \bibinfo{author}{\bibfnamefont{C.}~\bibnamefont{{Dominik}}},
  \bibinfo{author}{\bibfnamefont{F.}~\bibnamefont{{Meru}}},
  \bibinfo{author}{\bibfnamefont{M.~R.} \bibnamefont{{Meyer}}},
  \bibinfo{author}{\bibfnamefont{P.}~\bibnamefont{{Pinilla}}},
  \bibinfo{author}{\bibfnamefont{H.~M.} \bibnamefont{{Schmid}}},
  \bibnamefont{and} \bibinfo{author}{\bibfnamefont{S.}~\bibnamefont{{Wolf}}},
  \bibinfo{journal}{\aap} \textbf{\bibinfo{volume}{560}}, \bibinfo{eid}{A105}
  (\bibinfo{year}{2013}).

\bibitem[{\citenamefont{{Gautier} et~al.}(2012)\citenamefont{{Gautier},
  {Charbonneau}, {Rowe}, {Marcy}, {Isaacson}, {Torres}, {Fressin}, {Rogers},
  {D{\'e}sert}, {Buchhave} et~al.}}]{Gautier_Kepler20}
\bibinfo{author}{\bibfnamefont{T.~N.} \bibnamefont{{Gautier}},
  \bibfnamefont{III}},
  \bibinfo{author}{\bibfnamefont{D.}~\bibnamefont{{Charbonneau}}},
  \bibinfo{author}{\bibfnamefont{J.~F.} \bibnamefont{{Rowe}}},
  \bibinfo{author}{\bibfnamefont{G.~W.} \bibnamefont{{Marcy}}},
  \bibinfo{author}{\bibfnamefont{H.}~\bibnamefont{{Isaacson}}},
  \bibinfo{author}{\bibfnamefont{G.}~\bibnamefont{{Torres}}},
  \bibinfo{author}{\bibfnamefont{F.}~\bibnamefont{{Fressin}}},
  \bibinfo{author}{\bibfnamefont{L.~A.} \bibnamefont{{Rogers}}},
  \bibinfo{author}{\bibfnamefont{J.-M.} \bibnamefont{{D{\'e}sert}}},
  \bibinfo{author}{\bibfnamefont{L.~A.} \bibnamefont{{Buchhave}}},
  \bibnamefont{et~al.}, \bibinfo{journal}{\apj} \textbf{\bibinfo{volume}{749}},
  \bibinfo{eid}{15} (\bibinfo{year}{2012}), \eprint{1112.4514}.

\bibitem[{\citenamefont{{Haghighipour} and
  {Boss}}(2003)}]{Haghighipour_P_maxima}
\bibinfo{author}{\bibfnamefont{N.}~\bibnamefont{{Haghighipour}}}
  \bibnamefont{and} \bibinfo{author}{\bibfnamefont{A.~P.}
  \bibnamefont{{Boss}}}, \bibinfo{journal}{\apj}
  \textbf{\bibinfo{volume}{598}}, \bibinfo{pages}{1301} (\bibinfo{year}{2003}),
  \eprint{astro-ph/0305594}.

\bibitem[{\citenamefont{{Hsieh} and {Gu}}(2012)}]{Hsieh_Gu_dust_excitation}
\bibinfo{author}{\bibfnamefont{H.-F.} \bibnamefont{{Hsieh}}} \bibnamefont{and}
  \bibinfo{author}{\bibfnamefont{P.-G.} \bibnamefont{{Gu}}},
  \bibinfo{journal}{\apj} \textbf{\bibinfo{volume}{760}}, \bibinfo{eid}{119}
  (\bibinfo{year}{2012}), \eprint{1210.1648}.

\bibitem[{\citenamefont{{Isella} et~al.}(2013)\citenamefont{{Isella},
  {P{\'e}rez}, {Carpenter}, {Ricci}, {Andrews}, and
  {Rosenfeld}}}]{Isella_LkHalpha_multiplanet}
\bibinfo{author}{\bibfnamefont{A.}~\bibnamefont{{Isella}}},
  \bibinfo{author}{\bibfnamefont{L.~M.} \bibnamefont{{P{\'e}rez}}},
  \bibinfo{author}{\bibfnamefont{J.~M.} \bibnamefont{{Carpenter}}},
  \bibinfo{author}{\bibfnamefont{L.}~\bibnamefont{{Ricci}}},
  \bibinfo{author}{\bibfnamefont{S.}~\bibnamefont{{Andrews}}},
  \bibnamefont{and}
  \bibinfo{author}{\bibfnamefont{K.}~\bibnamefont{{Rosenfeld}}},
  \bibinfo{journal}{\apj} \textbf{\bibinfo{volume}{775}}, \bibinfo{eid}{30}
  (\bibinfo{year}{2013}), \eprint{1307.5848}.

\bibitem[{\citenamefont{{Klahr} and
  {Lin}}(2005)}]{Klahr_Lin_clumping_instability}
\bibinfo{author}{\bibfnamefont{H.}~\bibnamefont{{Klahr}}} \bibnamefont{and}
  \bibinfo{author}{\bibfnamefont{D.~N.~C.} \bibnamefont{{Lin}}},
  \bibinfo{journal}{\apj} \textbf{\bibinfo{volume}{632}}, \bibinfo{pages}{1113}
  (\bibinfo{year}{2005}), \eprint{astro-ph/0502536}.

\bibitem[{\citenamefont{{Kley}}(2000)}]{Kley_multiplanet}
\bibinfo{author}{\bibfnamefont{W.}~\bibnamefont{{Kley}}},
  \bibinfo{journal}{\mnras} \textbf{\bibinfo{volume}{313}},
  \bibinfo{pages}{L47} (\bibinfo{year}{2000}), \eprint{astro-ph/9910155}.

\bibitem[{\citenamefont{{Kraus} and {Ireland}}(2012)}]{LkCa15_protoplanet}
\bibinfo{author}{\bibfnamefont{A.~L.} \bibnamefont{{Kraus}}} \bibnamefont{and}
  \bibinfo{author}{\bibfnamefont{M.~J.} \bibnamefont{{Ireland}}},
  \bibinfo{journal}{\apj} \textbf{\bibinfo{volume}{745}}, \bibinfo{eid}{5}
  (\bibinfo{year}{2012}), \eprint{1110.3808}.

\bibitem[{\citenamefont{{Lin} and {Papaloizou}}(1993)}]{Lin_Papaloizou_PPIII}
\bibinfo{author}{\bibfnamefont{D.~N.~C.} \bibnamefont{{Lin}}} \bibnamefont{and}
  \bibinfo{author}{\bibfnamefont{J.~C.~B.} \bibnamefont{{Papaloizou}}}, in
  \emph{\bibinfo{booktitle}{Protostars and Planets III}}, edited by
  \bibinfo{editor}{\bibfnamefont{E.~H.} \bibnamefont{{Levy}}} \bibnamefont{and}
  \bibinfo{editor}{\bibfnamefont{J.~I.} \bibnamefont{{Lunine}}}
  (\bibinfo{year}{1993}), pp. \bibinfo{pages}{749--835}.

\bibitem[{\citenamefont{{Lyra} and
  {Kuchner}}(2013)}]{Lyra_Kuchner_clumping_instability}
\bibinfo{author}{\bibfnamefont{W.}~\bibnamefont{{Lyra}}} \bibnamefont{and}
  \bibinfo{author}{\bibfnamefont{M.}~\bibnamefont{{Kuchner}}},
  \bibinfo{journal}{\nat} \textbf{\bibinfo{volume}{499}}, \bibinfo{pages}{184}
  (\bibinfo{year}{2013}), \eprint{1307.5916}.

\bibitem[{\citenamefont{{Macintosh} et~al.}(2006)\citenamefont{{Macintosh},
  {Graham}, {Palmer}, {Doyon}, {Gavel}, {Larkin}, {Oppenheimer}, {Saddlemyer},
  {Wallace}, {Bauman} et~al.}}]{Macintosh_GPI}
\bibinfo{author}{\bibfnamefont{B.}~\bibnamefont{{Macintosh}}},
  \bibinfo{author}{\bibfnamefont{J.}~\bibnamefont{{Graham}}},
  \bibinfo{author}{\bibfnamefont{D.}~\bibnamefont{{Palmer}}},
  \bibinfo{author}{\bibfnamefont{R.}~\bibnamefont{{Doyon}}},
  \bibinfo{author}{\bibfnamefont{D.}~\bibnamefont{{Gavel}}},
  \bibinfo{author}{\bibfnamefont{J.}~\bibnamefont{{Larkin}}},
  \bibinfo{author}{\bibfnamefont{B.}~\bibnamefont{{Oppenheimer}}},
  \bibinfo{author}{\bibfnamefont{L.}~\bibnamefont{{Saddlemyer}}},
  \bibinfo{author}{\bibfnamefont{J.~K.} \bibnamefont{{Wallace}}},
  \bibinfo{author}{\bibfnamefont{B.}~\bibnamefont{{Bauman}}},
  \bibnamefont{et~al.}, in \emph{\bibinfo{booktitle}{Society of Photo-Optical
  Instrumentation Engineers (SPIE) Conference Series}} (\bibinfo{year}{2006}),
  vol. \bibinfo{volume}{6272} of \emph{\bibinfo{series}{Society of
  Photo-Optical Instrumentation Engineers (SPIE) Conference Series}}.

\bibitem[{\citenamefont{{Maddison} et~al.}(2007)\citenamefont{{Maddison},
  {Fouchet}, and {Gonzalez}}}]{Maddison_Neptune_dust_gap}
\bibinfo{author}{\bibfnamefont{S.~T.} \bibnamefont{{Maddison}}},
  \bibinfo{author}{\bibfnamefont{L.}~\bibnamefont{{Fouchet}}},
  \bibnamefont{and} \bibinfo{author}{\bibfnamefont{J.-F.}
  \bibnamefont{{Gonzalez}}}, \bibinfo{journal}{\apss}
  \textbf{\bibinfo{volume}{311}}, \bibinfo{pages}{3} (\bibinfo{year}{2007}),
  \eprint{0706.4248}.

\bibitem[{\citenamefont{{Malik} et~al.}(submitted)\citenamefont{{Malik},
  {Meru}, {Mayer}, and {Meyer}}}]{Malik_Meru_a}
\bibinfo{author}{\bibfnamefont{M.}~\bibnamefont{{Malik}}},
  \bibinfo{author}{\bibfnamefont{F.}~\bibnamefont{{Meru}}},
  \bibinfo{author}{\bibfnamefont{L.}~\bibnamefont{{Mayer}}}, \bibnamefont{and}
  \bibinfo{author}{\bibfnamefont{M.~R.} \bibnamefont{{Meyer}}},
  \bibinfo{journal}{}  (\bibinfo{year}{submitted}).

\bibitem[{\citenamefont{{Masset}}(2000)}]{Masset_FARGO}
\bibinfo{author}{\bibfnamefont{F.}~\bibnamefont{{Masset}}},
  \bibinfo{journal}{\aaps} \textbf{\bibinfo{volume}{141}}, \bibinfo{pages}{165}
  (\bibinfo{year}{2000}), \eprint{arXiv:astro-ph/9910390}.

\bibitem[{\citenamefont{{Masset} et~al.}(2006)\citenamefont{{Masset},
  {D'Angelo}, and {Kley}}}]{Masset_2.5RH}
\bibinfo{author}{\bibfnamefont{F.~S.} \bibnamefont{{Masset}}},
  \bibinfo{author}{\bibfnamefont{G.}~\bibnamefont{{D'Angelo}}},
  \bibnamefont{and} \bibinfo{author}{\bibfnamefont{W.}~\bibnamefont{{Kley}}},
  \bibinfo{journal}{\apj} \textbf{\bibinfo{volume}{652}}, \bibinfo{pages}{730}
  (\bibinfo{year}{2006}), \eprint{astro-ph/0607155}.

\bibitem[{\citenamefont{{Meru} et~al.}(2013)\citenamefont{{Meru},
  {Geretshauser}, {Sch{\"a}fer}, {Speith}, and {Kley}}}]{Velocity_thresholds}
\bibinfo{author}{\bibfnamefont{F.}~\bibnamefont{{Meru}}},
  \bibinfo{author}{\bibfnamefont{R.~J.} \bibnamefont{{Geretshauser}}},
  \bibinfo{author}{\bibfnamefont{C.}~\bibnamefont{{Sch{\"a}fer}}},
  \bibinfo{author}{\bibfnamefont{R.}~\bibnamefont{{Speith}}}, \bibnamefont{and}
  \bibinfo{author}{\bibfnamefont{W.}~\bibnamefont{{Kley}}},
  \bibinfo{journal}{\mnras}  (\bibinfo{year}{2013}), \eprint{1308.0825}.

\bibitem[{\citenamefont{{Okuzumi} et~al.}(2012)\citenamefont{{Okuzumi},
  {Tanaka}, {Kobayashi}, and {Wada}}}]{Okuzumi_RD_growth_Stokes}
\bibinfo{author}{\bibfnamefont{S.}~\bibnamefont{{Okuzumi}}},
  \bibinfo{author}{\bibfnamefont{H.}~\bibnamefont{{Tanaka}}},
  \bibinfo{author}{\bibfnamefont{H.}~\bibnamefont{{Kobayashi}}},
  \bibnamefont{and} \bibinfo{author}{\bibfnamefont{K.}~\bibnamefont{{Wada}}},
  \bibinfo{journal}{\apj} \textbf{\bibinfo{volume}{752}}, \bibinfo{eid}{106}
  (\bibinfo{year}{2012}), \eprint{1204.5035}.

\bibitem[{\citenamefont{{Osorio} et~al.}(2014)\citenamefont{{Osorio},
  {Anglada}, {Carrasco-Gonz{\'a}lez}, {Torrelles}, {Mac{\'{\i}}as},
  {Rodr{\'{\i}}guez}, {G{\'o}mez}, {D'Alessio}, {Calvet}, {Nagel}
  et~al.}}]{Osorio_HD169142_7mm}
\bibinfo{author}{\bibfnamefont{M.}~\bibnamefont{{Osorio}}},
  \bibinfo{author}{\bibfnamefont{G.}~\bibnamefont{{Anglada}}},
  \bibinfo{author}{\bibfnamefont{C.}~\bibnamefont{{Carrasco-Gonz{\'a}lez}}},
  \bibinfo{author}{\bibfnamefont{J.~M.} \bibnamefont{{Torrelles}}},
  \bibinfo{author}{\bibfnamefont{E.}~\bibnamefont{{Mac{\'{\i}}as}}},
  \bibinfo{author}{\bibfnamefont{L.~F.} \bibnamefont{{Rodr{\'{\i}}guez}}},
  \bibinfo{author}{\bibfnamefont{J.~F.} \bibnamefont{{G{\'o}mez}}},
  \bibinfo{author}{\bibfnamefont{P.}~\bibnamefont{{D'Alessio}}},
  \bibinfo{author}{\bibfnamefont{N.}~\bibnamefont{{Calvet}}},
  \bibinfo{author}{\bibfnamefont{E.}~\bibnamefont{{Nagel}}},
  \bibnamefont{et~al.}, \bibinfo{journal}{\apjl}
  \textbf{\bibinfo{volume}{791}}, \bibinfo{eid}{L36} (\bibinfo{year}{2014}),
  \eprint{1407.6549}.

\bibitem[{\citenamefont{{Paardekooper} and
  {Mellema}}(2004)}]{Paardekooper2004_dust}
\bibinfo{author}{\bibfnamefont{S.-J.} \bibnamefont{{Paardekooper}}}
  \bibnamefont{and}
  \bibinfo{author}{\bibfnamefont{G.}~\bibnamefont{{Mellema}}},
  \bibinfo{journal}{\aap} \textbf{\bibinfo{volume}{425}}, \bibinfo{pages}{L9}
  (\bibinfo{year}{2004}), \eprint{astro-ph/0408202}.

\bibitem[{\citenamefont{{Paardekooper} and
  {Mellema}}(2006)}]{Paardekooper2006_dust}
\bibinfo{author}{\bibfnamefont{S.-J.} \bibnamefont{{Paardekooper}}}
  \bibnamefont{and}
  \bibinfo{author}{\bibfnamefont{G.}~\bibnamefont{{Mellema}}},
  \bibinfo{journal}{\aap} \textbf{\bibinfo{volume}{453}}, \bibinfo{pages}{1129}
  (\bibinfo{year}{2006}), \eprint{astro-ph/0603132}.

\bibitem[{\citenamefont{{P{\'e}rez} et~al.}(2014)\citenamefont{{P{\'e}rez},
  {Isella}, {Carpenter}, and {Chandler}}}]{Perez_TD_asymmetries}
\bibinfo{author}{\bibfnamefont{L.~M.} \bibnamefont{{P{\'e}rez}}},
  \bibinfo{author}{\bibfnamefont{A.}~\bibnamefont{{Isella}}},
  \bibinfo{author}{\bibfnamefont{J.~M.} \bibnamefont{{Carpenter}}},
  \bibnamefont{and} \bibinfo{author}{\bibfnamefont{C.~J.}
  \bibnamefont{{Chandler}}}, \bibinfo{journal}{\apjl}
  \textbf{\bibinfo{volume}{783}}, \bibinfo{eid}{L13} (\bibinfo{year}{2014}),
  \eprint{1402.0832}.

\bibitem[{\citenamefont{{Pierens} and
  {Nelson}}(2008)}]{Pierens_Nelson_2planets}
\bibinfo{author}{\bibfnamefont{A.}~\bibnamefont{{Pierens}}} \bibnamefont{and}
  \bibinfo{author}{\bibfnamefont{R.~P.} \bibnamefont{{Nelson}}},
  \bibinfo{journal}{\aap} \textbf{\bibinfo{volume}{482}}, \bibinfo{pages}{333}
  (\bibinfo{year}{2008}), \eprint{0802.2033}.

\bibitem[{\citenamefont{{Pineda} et~al.}(2014)\citenamefont{{Pineda}, {Quanz},
  {Meru}, {Mulders}, {Meyer}, {Pani{\'c}}, and {Avenhaus}}}]{Pineda_HD100546}
\bibinfo{author}{\bibfnamefont{J.~E.} \bibnamefont{{Pineda}}},
  \bibinfo{author}{\bibfnamefont{S.~P.} \bibnamefont{{Quanz}}},
  \bibinfo{author}{\bibfnamefont{F.}~\bibnamefont{{Meru}}},
  \bibinfo{author}{\bibfnamefont{G.~D.} \bibnamefont{{Mulders}}},
  \bibinfo{author}{\bibfnamefont{M.~R.} \bibnamefont{{Meyer}}},
  \bibinfo{author}{\bibfnamefont{O.}~\bibnamefont{{Pani{\'c}}}},
  \bibnamefont{and}
  \bibinfo{author}{\bibfnamefont{H.}~\bibnamefont{{Avenhaus}}},
  \bibinfo{journal}{\apjl} \textbf{\bibinfo{volume}{788}}, \bibinfo{eid}{L34}
  (\bibinfo{year}{2014}), \eprint{1405.5773}.

\bibitem[{\citenamefont{{Pinilla}
  et~al.}(2012{\natexlab{b}})\citenamefont{{Pinilla}, {Birnstiel}, {Ricci},
  {Dullemond}, {Uribe}, {Testi}, and {Natta}}}]{Pinilla_pressure_bump}
\bibinfo{author}{\bibfnamefont{P.}~\bibnamefont{{Pinilla}}},
  \bibinfo{author}{\bibfnamefont{T.}~\bibnamefont{{Birnstiel}}},
  \bibinfo{author}{\bibfnamefont{L.}~\bibnamefont{{Ricci}}},
  \bibinfo{author}{\bibfnamefont{C.~P.} \bibnamefont{{Dullemond}}},
  \bibinfo{author}{\bibfnamefont{A.~L.} \bibnamefont{{Uribe}}},
  \bibinfo{author}{\bibfnamefont{L.}~\bibnamefont{{Testi}}}, \bibnamefont{and}
  \bibinfo{author}{\bibfnamefont{A.}~\bibnamefont{{Natta}}},
  \bibinfo{journal}{\aap} \textbf{\bibinfo{volume}{538}}, \bibinfo{eid}{A114}
  (\bibinfo{year}{2012}{\natexlab{b}}), \eprint{1112.2349}.

\bibitem[{\citenamefont{{Pinilla}
  et~al.}(2012{\natexlab{a}})\citenamefont{{Pinilla}, {Benisty}, and
  {Birnstiel}}}]{Pinilla_planet_dust}
\bibinfo{author}{\bibfnamefont{P.}~\bibnamefont{{Pinilla}}},
  \bibinfo{author}{\bibfnamefont{M.}~\bibnamefont{{Benisty}}},
  \bibnamefont{and}
  \bibinfo{author}{\bibfnamefont{T.}~\bibnamefont{{Birnstiel}}},
  \bibinfo{journal}{\aap} \textbf{\bibinfo{volume}{545}}, \bibinfo{eid}{A81}
  (\bibinfo{year}{2012}{\natexlab{a}}), \eprint{1207.6485}.

\bibitem[{\citenamefont{{Pinilla} et~al.}(2014)\citenamefont{{Pinilla}, {de
  Juan Ovelar}, {Ataiee}, {Benisty}, {Birnstiel}, {van Dishoeck}, and
  {Min}}}]{Pinilla_multiplanet}
\bibinfo{author}{\bibfnamefont{P.}~\bibnamefont{{Pinilla}}},
  \bibinfo{author}{\bibfnamefont{M.}~\bibnamefont{{de Juan Ovelar}}},
  \bibinfo{author}{\bibfnamefont{S.}~\bibnamefont{{Ataiee}}},
  \bibinfo{author}{\bibfnamefont{M.}~\bibnamefont{{Benisty}}},
  \bibinfo{author}{\bibfnamefont{T.}~\bibnamefont{{Birnstiel}}},
  \bibinfo{author}{\bibfnamefont{E.~F.} \bibnamefont{{van Dishoeck}}},
  \bibnamefont{and} \bibinfo{author}{\bibfnamefont{M.}~\bibnamefont{{Min}}},
  \bibinfo{journal}{ArXiv e-prints}  (\bibinfo{year}{2014}),
  \eprint{1410.5963}.

\bibitem[{\citenamefont{{Quanz}
  et~al.}(2013{\natexlab{a}})\citenamefont{{Quanz}, {Avenhaus}, {Buenzli},
  {Garufi}, {Schmid}, and {Wolf}}}]{Quanz_doughnut}
\bibinfo{author}{\bibfnamefont{S.~P.} \bibnamefont{{Quanz}}},
  \bibinfo{author}{\bibfnamefont{H.}~\bibnamefont{{Avenhaus}}},
  \bibinfo{author}{\bibfnamefont{E.}~\bibnamefont{{Buenzli}}},
  \bibinfo{author}{\bibfnamefont{A.}~\bibnamefont{{Garufi}}},
  \bibinfo{author}{\bibfnamefont{H.~M.} \bibnamefont{{Schmid}}},
  \bibnamefont{and} \bibinfo{author}{\bibfnamefont{S.}~\bibnamefont{{Wolf}}},
  \bibinfo{journal}{\apjl} \textbf{\bibinfo{volume}{766}}, \bibinfo{eid}{L2}
  (\bibinfo{year}{2013}{\natexlab{a}}), \eprint{1302.3029}.

\bibitem[{\citenamefont{{Quanz}
  et~al.}(2013{\natexlab{b}})\citenamefont{{Quanz}, {Amara}, {Meyer},
  {Kenworthy}, {Kasper}, and {Girard}}}]{HD100546b}
\bibinfo{author}{\bibfnamefont{S.~P.} \bibnamefont{{Quanz}}},
  \bibinfo{author}{\bibfnamefont{A.}~\bibnamefont{{Amara}}},
  \bibinfo{author}{\bibfnamefont{M.~R.} \bibnamefont{{Meyer}}},
  \bibinfo{author}{\bibfnamefont{M.~A.} \bibnamefont{{Kenworthy}}},
  \bibinfo{author}{\bibfnamefont{M.}~\bibnamefont{{Kasper}}}, \bibnamefont{and}
  \bibinfo{author}{\bibfnamefont{J.~H.} \bibnamefont{{Girard}}},
  \bibinfo{journal}{\apjl} \textbf{\bibinfo{volume}{766}}, \bibinfo{eid}{L1}
  (\bibinfo{year}{2013}{\natexlab{b}}), \eprint{1302.7122}.

\bibitem[{\citenamefont{{Quanz et
  al}}(submitted)}]{Quanz_HD100546_confirmation}
\bibinfo{author}{\bibfnamefont{S.~P.} \bibnamefont{{Quanz et al}}}
  (\bibinfo{year}{submitted}).

\bibitem[{\citenamefont{{Reggiani} et~al.}(2014)\citenamefont{{Reggiani},
  {Quanz}, {Meyer}, {Pueyo}, {Absil}, {Amara}, {Anglada}, {Avenhaus}, {Girard},
  {Carrasco Gonzalez} et~al.}}]{Reggiani_HD169142}
\bibinfo{author}{\bibfnamefont{M.}~\bibnamefont{{Reggiani}}},
  \bibinfo{author}{\bibfnamefont{S.~P.} \bibnamefont{{Quanz}}},
  \bibinfo{author}{\bibfnamefont{M.~R.} \bibnamefont{{Meyer}}},
  \bibinfo{author}{\bibfnamefont{L.}~\bibnamefont{{Pueyo}}},
  \bibinfo{author}{\bibfnamefont{O.}~\bibnamefont{{Absil}}},
  \bibinfo{author}{\bibfnamefont{A.}~\bibnamefont{{Amara}}},
  \bibinfo{author}{\bibfnamefont{G.}~\bibnamefont{{Anglada}}},
  \bibinfo{author}{\bibfnamefont{H.}~\bibnamefont{{Avenhaus}}},
  \bibinfo{author}{\bibfnamefont{J.~H.} \bibnamefont{{Girard}}},
  \bibinfo{author}{\bibfnamefont{C.}~\bibnamefont{{Carrasco Gonzalez}}},
  \bibnamefont{et~al.}, \bibinfo{journal}{ArXiv e-prints}
  (\bibinfo{year}{2014}), \eprint{1408.0813}.

\bibitem[{\citenamefont{{Schmid} et~al.}(2010)\citenamefont{{Schmid}, {Beuzit},
  {Mouillet}, {Waters}, {Buenzli}, {Boccaletti}, {Dohlen}, {Feldt}, and {SPHERE
  Consortium}}}]{schmid2006}
\bibinfo{author}{\bibfnamefont{H.~M.} \bibnamefont{{Schmid}}},
  \bibinfo{author}{\bibfnamefont{J.~L.} \bibnamefont{{Beuzit}}},
  \bibinfo{author}{\bibfnamefont{D.}~\bibnamefont{{Mouillet}}},
  \bibinfo{author}{\bibfnamefont{R.}~\bibnamefont{{Waters}}},
  \bibinfo{author}{\bibfnamefont{E.}~\bibnamefont{{Buenzli}}},
  \bibinfo{author}{\bibfnamefont{A.}~\bibnamefont{{Boccaletti}}},
  \bibinfo{author}{\bibfnamefont{K.}~\bibnamefont{{Dohlen}}},
  \bibinfo{author}{\bibfnamefont{M.}~\bibnamefont{{Feldt}}}, \bibnamefont{and}
  \bibinfo{author}{\bibnamefont{{SPHERE Consortium}}}, in
  \emph{\bibinfo{booktitle}{In the Spirit of Lyot 2010}}
  (\bibinfo{year}{2010}), p.~\bibinfo{pages}{49}.

\bibitem[{\citenamefont{{Teiser} and {Wurm}}(2009)}]{Teiser_Wurm_highVcoll}
\bibinfo{author}{\bibfnamefont{J.}~\bibnamefont{{Teiser}}} \bibnamefont{and}
  \bibinfo{author}{\bibfnamefont{G.}~\bibnamefont{{Wurm}}},
  \bibinfo{journal}{\mnras} \textbf{\bibinfo{volume}{393}},
  \bibinfo{pages}{1584} (\bibinfo{year}{2009}), \eprint{0901.4235}.

\bibitem[{\citenamefont{{Thalmann} et~al.}(2014)\citenamefont{{Thalmann},
  {Mulders}, {Hodapp}, {Janson}, {Grady}, {Min}, {de Juan Ovelar}, {Carson},
  {Brandt}, {Bonnefoy} et~al.}}]{Thalmann_LkCa15_ecc}
\bibinfo{author}{\bibfnamefont{C.}~\bibnamefont{{Thalmann}}},
  \bibinfo{author}{\bibfnamefont{G.~D.} \bibnamefont{{Mulders}}},
  \bibinfo{author}{\bibfnamefont{K.}~\bibnamefont{{Hodapp}}},
  \bibinfo{author}{\bibfnamefont{M.}~\bibnamefont{{Janson}}},
  \bibinfo{author}{\bibfnamefont{C.~A.} \bibnamefont{{Grady}}},
  \bibinfo{author}{\bibfnamefont{M.}~\bibnamefont{{Min}}},
  \bibinfo{author}{\bibfnamefont{M.}~\bibnamefont{{de Juan Ovelar}}},
  \bibinfo{author}{\bibfnamefont{J.}~\bibnamefont{{Carson}}},
  \bibinfo{author}{\bibfnamefont{T.}~\bibnamefont{{Brandt}}},
  \bibinfo{author}{\bibfnamefont{M.}~\bibnamefont{{Bonnefoy}}},
  \bibnamefont{et~al.}, \bibinfo{journal}{\aap} \textbf{\bibinfo{volume}{566}},
  \bibinfo{eid}{A51} (\bibinfo{year}{2014}), \eprint{1402.1766}.

\bibitem[{\citenamefont{{Tremaine}}(1998)}]{Tremaine_secular}
\bibinfo{author}{\bibfnamefont{S.}~\bibnamefont{{Tremaine}}},
  \bibinfo{journal}{\aj} \textbf{\bibinfo{volume}{116}}, \bibinfo{pages}{2015}
  (\bibinfo{year}{1998}), \eprint{astro-ph/9805334}.

\bibitem[{\citenamefont{{van der Marel} et~al.}(2013)\citenamefont{{van der
  Marel}, {van Dishoeck}, {Bruderer}, {Birnstiel}, {Pinilla}, {Dullemond}, {van
  Kempen}, {Schmalzl}, {Brown}, {Herczeg} et~al.}}]{Nienke_dust_trap}
\bibinfo{author}{\bibfnamefont{N.}~\bibnamefont{{van der Marel}}},
  \bibinfo{author}{\bibfnamefont{E.~F.} \bibnamefont{{van Dishoeck}}},
  \bibinfo{author}{\bibfnamefont{S.}~\bibnamefont{{Bruderer}}},
  \bibinfo{author}{\bibfnamefont{T.}~\bibnamefont{{Birnstiel}}},
  \bibinfo{author}{\bibfnamefont{P.}~\bibnamefont{{Pinilla}}},
  \bibinfo{author}{\bibfnamefont{C.~P.} \bibnamefont{{Dullemond}}},
  \bibinfo{author}{\bibfnamefont{T.~A.} \bibnamefont{{van Kempen}}},
  \bibinfo{author}{\bibfnamefont{M.}~\bibnamefont{{Schmalzl}}},
  \bibinfo{author}{\bibfnamefont{J.~M.} \bibnamefont{{Brown}}},
  \bibinfo{author}{\bibfnamefont{G.~J.} \bibnamefont{{Herczeg}}},
  \bibnamefont{et~al.}, \bibinfo{journal}{Science}
  \textbf{\bibinfo{volume}{340}}, \bibinfo{pages}{1199} (\bibinfo{year}{2013}),
  \eprint{1306.1768}.

\bibitem[{\citenamefont{{Walsh} et~al.}(2014)\citenamefont{{Walsh},
  {Juh{\'a}sz}, {Pinilla}, {Harsono}, {Mathews}, {Dent}, {Hogerheijde},
  {Birnstiel}, {Meeus}, {Nomura} et~al.}}]{Walsh_HD100546}
\bibinfo{author}{\bibfnamefont{C.}~\bibnamefont{{Walsh}}},
  \bibinfo{author}{\bibfnamefont{A.}~\bibnamefont{{Juh{\'a}sz}}},
  \bibinfo{author}{\bibfnamefont{P.}~\bibnamefont{{Pinilla}}},
  \bibinfo{author}{\bibfnamefont{D.}~\bibnamefont{{Harsono}}},
  \bibinfo{author}{\bibfnamefont{G.~S.} \bibnamefont{{Mathews}}},
  \bibinfo{author}{\bibfnamefont{W.~R.~F.} \bibnamefont{{Dent}}},
  \bibinfo{author}{\bibfnamefont{M.~R.} \bibnamefont{{Hogerheijde}}},
  \bibinfo{author}{\bibfnamefont{T.}~\bibnamefont{{Birnstiel}}},
  \bibinfo{author}{\bibfnamefont{G.}~\bibnamefont{{Meeus}}},
  \bibinfo{author}{\bibfnamefont{H.}~\bibnamefont{{Nomura}}},
  \bibnamefont{et~al.}, \bibinfo{journal}{\apjl}
  \textbf{\bibinfo{volume}{791}}, \bibinfo{eid}{L6} (\bibinfo{year}{2014}),
  \eprint{1405.6542}.

\bibitem[{\citenamefont{{Weidenschilling}}(1977)}]{Weidenschilling1977}
\bibinfo{author}{\bibfnamefont{S.~J.} \bibnamefont{{Weidenschilling}}},
  \bibinfo{journal}{\mnras} \textbf{\bibinfo{volume}{180}}, \bibinfo{pages}{57}
  (\bibinfo{year}{1977}).

\bibitem[{\citenamefont{{Wurm} et~al.}(2005)\citenamefont{{Wurm}, {Paraskov},
  and {Krauss}}}]{Wurm_25m/s_impacts}
\bibinfo{author}{\bibfnamefont{G.}~\bibnamefont{{Wurm}}},
  \bibinfo{author}{\bibfnamefont{G.}~\bibnamefont{{Paraskov}}},
  \bibnamefont{and} \bibinfo{author}{\bibfnamefont{O.}~\bibnamefont{{Krauss}}},
  \bibinfo{journal}{\icarus} \textbf{\bibinfo{volume}{178}},
  \bibinfo{pages}{253} (\bibinfo{year}{2005}).

\bibitem[{\citenamefont{{Zhu} et~al.}(2011)\citenamefont{{Zhu}, {Nelson},
  {Hartmann}, {Espaillat}, and {Calvet}}}]{Zhu_multiplanet}
\bibinfo{author}{\bibfnamefont{Z.}~\bibnamefont{{Zhu}}},
  \bibinfo{author}{\bibfnamefont{R.~P.} \bibnamefont{{Nelson}}},
  \bibinfo{author}{\bibfnamefont{L.}~\bibnamefont{{Hartmann}}},
  \bibinfo{author}{\bibfnamefont{C.}~\bibnamefont{{Espaillat}}},
  \bibnamefont{and} \bibinfo{author}{\bibfnamefont{N.}~\bibnamefont{{Calvet}}},
  \bibinfo{journal}{\apj} \textbf{\bibinfo{volume}{729}}, \bibinfo{eid}{47}
  (\bibinfo{year}{2011}), \eprint{1012.4395}.

\bibitem[{\citenamefont{{Zhu} et~al.}(2012)\citenamefont{{Zhu}, {Nelson},
  {Dong}, {Espaillat}, and {Hartmann}}}]{Zhu_dust_filtration}
\bibinfo{author}{\bibfnamefont{Z.}~\bibnamefont{{Zhu}}},
  \bibinfo{author}{\bibfnamefont{R.~P.} \bibnamefont{{Nelson}}},
  \bibinfo{author}{\bibfnamefont{R.}~\bibnamefont{{Dong}}},
  \bibinfo{author}{\bibfnamefont{C.}~\bibnamefont{{Espaillat}}},
  \bibnamefont{and}
  \bibinfo{author}{\bibfnamefont{L.}~\bibnamefont{{Hartmann}}},
  \bibinfo{journal}{\apj} \textbf{\bibinfo{volume}{755}}, \bibinfo{eid}{6}
  (\bibinfo{year}{2012}), \eprint{1205.5042}.

\bibitem[{\citenamefont{{Zhu} et~al.}(2014)\citenamefont{{Zhu}, {Stone},
  {Rafikov}, and {Bai}}}]{Zhu_2D_3D_dust_gaps}
\bibinfo{author}{\bibfnamefont{Z.}~\bibnamefont{{Zhu}}},
  \bibinfo{author}{\bibfnamefont{J.~M.} \bibnamefont{{Stone}}},
  \bibinfo{author}{\bibfnamefont{R.~R.} \bibnamefont{{Rafikov}}},
  \bibnamefont{and} \bibinfo{author}{\bibfnamefont{X.-n.} \bibnamefont{{Bai}}},
  \bibinfo{journal}{\apj} \textbf{\bibinfo{volume}{785}}, \bibinfo{eid}{122}
  (\bibinfo{year}{2014}), \eprint{1308.0648}.


\end{thebibliography}
\end{document}